\documentclass[aps,prd,onecolumn,floatfix,superscriptaddress,nofootinbib]{revtex4-1}
\newcommand{\gev}{\ensuremath{\,\mathrm{GeV}}}
\newcommand{\tev}{\ensuremath{\,\mathrm{TeV}}}

\usepackage{amsmath}
\usepackage{graphicx,bm}
\usepackage[colorlinks,citecolor=blue]{hyperref}
\usepackage[caption=false]{subfig}
\usepackage{slashed}

\begin{document}
\title{Elastic and Inelastic Scattering of Cosmic-Rays on sub-GeV Dark Matter}

\author{Gang Guo}
\email{gangg23@gmail.com}
\affiliation{Institute of Physics, Academia Sinica, Taipei, 11529, Taiwan}

\author{Yue-Lin Sming Tsai}
\email{smingtsai@gate.sinica.edu.tw}
\affiliation{Institute of Physics, Academia Sinica, Taipei, 11529, Taiwan}
\affiliation{Key Laboratory of Dark Matter and Space Astronomy, Purple Mountain Observatory, Chinese Academy of Sciences, Nanjing 210033, China}
\affiliation{Department of Physics, National Tsing Hua University,
Hsinchu 300, Taiwan}

\author{Meng-Ru~Wu}
\email{mwu@gate.sinica.edu.tw}
\affiliation{Institute of Physics, Academia Sinica, Taipei, 11529, Taiwan}
\affiliation{Institute of Astronomy and Astrophysics, Academia Sinica, Taipei, 10617, Taiwan}
\affiliation{National Center for Theoretical Sciences, Physics Division, Hsinchu, 30013, Taiwan}

\author{Qiang Yuan}
\email{yuanq@pmo.ac.cn}
\affiliation{Key Laboratory of Dark Matter and Space Astronomy, Purple Mountain Observatory, Chinese Academy of Sciences, Nanjing 210033, China}
\affiliation{School of Astronomy and Space Science, University of Science and Technology of China, Hefei, Anhui 230026, China}
\affiliation{Center for High Energy Physics, Peking University, Beijing 100871, China}

%\affiliation[a]{Institute of Physics, Academia Sinica, Taipei, 11529, Taiwan}
%\affiliation[b]{Key Laboratory of Dark Matter and Space Astronomy, Purple Mountain Observatory, Chinese Academy of Sciences, Nanjing 210008, China}
%\affiliation[c]{Institute of Astronomy and Astrophysics, Academia Sinica, Taipei, 10617, Taiwan}
%\affiliation[d]{National Center for Theoretical Sciences, Physics Division, Hsinchu, 30013, Taiwan}
%\emailAdd{gangg23@gmail.com}
%\emailAdd{smingtsai@gate.sinica.edu.tw}
%\emailAdd{mwu@gate.sinica.edu.tw}

\begin{abstract}
We revisit the signatures from collisions of cosmic-rays on sub-GeV dark matter (DM) in the Milky Way. In addition to the upscattered DM component that can be probed by existing DM and neutrino experiments widely discussed, we examine the associated signals in $\gamma$-rays and neutrinos that span a wide energy range due to the inelastic scatterings. Assuming a simple vector portal DM model for illustration, 
we compute both the upscattered DM flux by cosmic-ray protons, and the
resulting emission of secondary $\gamma$-rays and high-energy neutrinos
from proton excitation, hadronization, and the subsequent meson decay. We derive limits on coupling constants in the vector portal model using data from the $\gamma$-ray and high-energy 
neutrino telescopes including Fermi, H.E.S.S. and IceCube. 
These limits are
compared to those obtained by considering the upscattered DM signals at the low-energy DM/neutrino detectors XENON1T/MiniBooNE and the IceCube.
For this particular model, the limits are set predominantly by non-detection of the upscattered DM events in XENON1T, for most of the DM mass range due to the large scattering cross section at low energies.
Nevertheless, our study demonstrates that the $\gamma$-ray and neutrino signals, traditionally considered as indirect probes for DM annihilation and decay, can also be directly used to constrain the DM--nucleon interaction in complementary to the direct search experiments.
\end{abstract}

\date{\today}

%\begin{document}
\maketitle
%\flushbottom

\section{introduction}

The extensive searches for
the interaction between particle dark matter (DM) and the standard model (SM) sectors are conventionally classified into three different categories: the direct detection via the scattering between the DM and SM particles (see, e.g., \cite{Akerib:2016vxi,Akerib:2017kat,Cui:2017nnn,Fu:2016ega,Aprile:2018dbl,Aprile:2019dbj} for the most sensitive searches for massive DM particles, and
for recent reviews see, e.g., \cite{Liu:2017drf,Schumann:2019eaa}), the indirect search for signals from DM annihilation or decay (see, e.g., \cite{Gaskins:2016cha,Leane:2020liq,PerezdelosHeros:2020qyt}), and the search of DM production in colliders (see, e.g., \cite{Buchmueller:2017qhf,Kahlhoefer:2017dnp,Boveia:2018yeb}).
Although they can, in principle, be used together to test a given model describing how DM interact with the SM sectors, the involved physical processes in generating the signals are clearly distinct in different categories. Despite great efforts to hunt for DM in all these respects during the past few decades, null signal of DM has been clearly confirmed to date. 
Usually, the search for DM in each category could be subject to different limitations. 
For example, the indirect searches focusing on signals from DM annihilation or decay 
cannot be used to probe asymmetric or stable DM. Moreover, the deep underground direct detection experiments lose sensitivity rapidly for sub-GeV DM with recoil energies below the detection threshold (see, e.g., \cite{Akerib:2016vxi,Akerib:2017kat,Cui:2017nnn,Fu:2016ega,Aprile:2018dbl,Aprile:2019dbj}).

New detection techniques or possibilities have been recently investigated to detect dark matter lighter than GeV (see, e.g., \cite{Knapen:2016cue,An:2017ojc,Ibe:2017yqa,Berlin:2018sjs,Akesson:2018vlm,Matsumoto:2018acr,Berlin:2019uco,Depta:2019lbe,Hertel:2019thc,Dror:2019onn,Dror:2019dib}). Among them, the idea of detecting a subdominant fraction of DM particles that are unavoidably upscattered by cosmic-rays (CRs) to high enough energies was recently proposed in Refs.~\cite{Bringmann:2018cvk,Ema:2018bih} and studied subsequently in Ref.~\cite{Cappiello:2019qsw}. Assuming an energy-independent cross section, the subsequent scattering of the accelerated DM inside low-energy DM and neutrino detectors can be used to derive new constraints on DM-nucleon or DM-electron cross sections for light DM. Ref.~\cite{Guo:2020drq} considered the case that DM can be upscattered to high-energies of above $\sim$PeV that can be detected by IceCube, providing a probe of DM-nucleon cross section at such high-energy (HE) scales. Relevant studies also investigated the diurnal effect of the boosted DM signals \cite{Ge:2020yuf} 
and those at germanium detectors \cite{Zhang:2020htl}.
In addition to all these model-independent studies above, Refs.~\cite{Bondarenko:2019vrb,Dent:2019krz,Wang:2019jtk,Cho:2020mnc,Cao:2020bwd} explored the CR-boosted DM signals based on simple microscopic DM models and placed limits on the coupling constants and masses related to the dark sectors. Very recently,
an excess of ${\cal O}$(keV) electronic recoil events has been observed in XENON1T \cite{Aprile:2020tmw}. The scenarios of CR-boosted DM have also been employed to explain such an anomaly \cite{Jho:2020sku,Bloch:2020uzh}. In relevance to the role of CRs,
similar constraints based on the distortion of the CR spectrum at $\sim$TeV due to scattering with DM~\cite{Cappiello:2018hsu} or the
DM signals produced from the collision of CRs with the atmosphere \cite{Alvey:2019zaa,Plestid:2020kdm,Su:2020zny} were also discussed.

We note that most of these studies on CR-boosted DM signals in the existing literature are limited to only elastic $\chi$--$p$ scattering. However, for CRs colliding with DM with a center-of-mass energy $\gtrsim$~GeV, the nucleons can be excited and, consequently, produce secondary $\gamma$-rays \cite{Cyburt:2002uw,Hooper:2018bfw} and neutrinos in the decay sequence.
Moreover, at higher energy scales, DM can scatter off individual quarks inside nucleons and lead to deep inelastic scatterings (DIS) that will also generate secondary neutrinos and $\gamma$-rays from the hadronization and the subsequent meson decay. These inelastic scatterings are just like the $pp$, $p\gamma$, or $p\nu$ collisions in the SM and are bound to occur regardless of the particle nature of DM. Consequently, the co-produced ``indirect'' signals testable by HE $\gamma$-ray and neutrino telescopes can simultaneously probe the DM--nucleon interaction together with the direct detection of the upscattered DM by CRs without invoking any other assumptions.

In this work, we study consistently upscattered DM by CRs, together with the co-produced secondary HE $\gamma$-ray and neutrino emissions for the first time.
For the purpose of including the inelastic scatterings, it is necessary to adopt a specific model describing the interaction between the DM and SM particles.
Especially, the excitation of nucleons to resonances with different spins or isospins depends explicitly on the interaction type.
More importantly,
as we aim to cover a wide energy scale over several orders of magnitudes in this study, relevant to low-energy elastic $\chi$-$p$ scattering as well as $\gamma$-ray and HE neutrino productions from inelastic scattering, a consistent treatment based on a specific model taking fully into account the energy dependence as done in Refs.~\cite{Bondarenko:2019vrb,Dent:2019krz,Wang:2019jtk,Cho:2020mnc} is required. This is different from the studies in Refs.~\cite{Cyburt:2002uw,Hooper:2018bfw} which adopted either a constant inelastic cross section for $\pi^0$ production \cite{Cyburt:2002uw} or a rescaled 
cross section following the energy dependence of $pp$ collision \cite{Hooper:2018bfw}. 
In addition, we also include $\gamma$-rays from other hadrons and consider constraints due to the secondary neutrino emission.
For illustration, we simply consider a fermionic DM which couples to baryons via a new dark vector boson. The corresponding Lagrangian is
\begin{equation}
\mathcal{L} \supset \overline{\chi}(i\partial_\mu\gamma^\mu-m_{\chi})\chi + 
g_\chi \overline{\chi} \gamma^\mu \chi V_\mu + \sum_{f=u,d,s,...} g_{q_f}\overline{q}_f \gamma^{\mu} q_f V_\mu 
+\frac{1}{2} m^2_V V_\mu V^{\mu}\,,
\label{eq:dmlag}
\end{equation}
where $\chi$, $V_\mu$, and $q_f$ refer to the DM, the vector mediator, and quarks, respectively. 
For simplicity, we consider only that the vector mediator couples to quarks, as only scatterings between DM and baryons are relevant for our study. 
Furthermore, we consider a universal coupling constant between the vector mediator and quarks of different flavors, i.e., $g_{q_f} = g_q = g_B/3$. 
This can be realized when the vector boson is a $U(1)_B$ gauge boson with $B$ the baryon number. We should point out that the simple $U(1)_B$ model considered in our work suffers from gauge anomalies. To realize a UV completion of the model, new heavy particles charged under the new gauge should be introduced to cancel the anomalies. However, searches for these heavy particles with collider experiments~\cite{Dobrescu:2014fca} as well as the decay of $Z$ and mesons~\cite{Dror:2017ehi} can place tight constraints on the viable parameter space. Alternative options are to consider the anomaly free models such as 
a gauged $B-L$ model (\cite{Foot:1990mn}) or a kinetically-mixed dark photon model \cite{Okun:1982xi,Holdom:1985ag}. These models with light gauged mediators are also strongly constrained by terrestrial experiments as well as astrophysical and cosmological observations (see, e.g., \cite{An:2013yfc,Batell:2014mga,Heeck:2014zfa,Bilmis:2015lja,Fayet:2016nyc,Bauer:2018onh,Chang:2018rso,Sung:2019xie,DeRocco:2019njg,Li:2020roy,Knapen:2017xzo,Alexander:2016aln} and a recent review on dark photon \cite{Fabbrichesi:2020wbt}). Even the gauged $U(1)_B$ model without considering the anomalies are tightly constrained by terrestrial experiments (see, e.g., \cite{Barbieri:1975xy,Batell:2014yra,Dror:2017nsg,Knapen:2017xzo,Aguilar-Arevalo:2017mqx,Aguilar-Arevalo:2018wea}), BBN observation \cite{Krnjaic:2019dzc} as well as supernova cooling \cite{Rrapaj:2015wgs}.
As mentioned above, in this work we simply take the $U(1)_B$ model as an example and demonstrate that all the relevant signatures of elastic/inelastic collisions between HE CRs and DM can be consistently studied, and can be used to constrain the DM models, even though the associated constraints could be weaker than those derived from other methods. To be more specific, we derive exclusion limits on
the couplings for sub-GeV DM by considering the upscattered DM component with XENON1T~\cite{Aprile:2015uzo,Aprile:2017aty}, MiniBooNE~\cite{AguilarArevalo:2008qa}, IceCube~\cite{Aartsen:2016nxy}, the secondary $\gamma$-rays with data from Fermi~\cite{TheFermi-LAT:2017vmf} and H.E.S.S.~\cite{Abramowski:2016mir}, as well as the secondary neutrinos with IceCube.
Although we reply on this specific model to explore the inelastic effects, the same method described
in this work can also be applied
to other DM models.

This paper is organized as follows. In Sec.~\ref{sec:xsec}, 
we explicitly compute the cross sections of elastic and inelastic scatterings 
between $\chi$ and $p$ based on the Lagrangian given in Eq.~\eqref{eq:dmlag}.
The upscattered DM flux by CRs in the Milky Way (MW)
as well as the fluxes of the secondary HE $\gamma$-rays and neutrinos are derived.
For demonstration, we show
our results based on two particular choices of vector boson mass:
(i) $m_V=3 m_\chi$ that is widely adopted in the literature (see, e.g., \cite{Chang:2018rso,Fabbrichesi:2020wbt,Aguilar-Arevalo:2017mqx,Aguilar-Arevalo:2018wea}) and 
for which dark bosons decaying to DM pair is kinematically allowed; (ii) a light mediator case with $m_V=1$ eV. \footnote{\label{ft1}We point out that our results in this work are actually insensitive to the value of $m_V$ if $m_V^2$ is much smaller than the typical value of $Q^2$ involved in DM-nucleon scatterings, where $Q^2$ is the four-momentum transfer squared. For example, $Q^2 \gtrsim (0.1~{\rm GeV})^2$ for $\gamma$-ray production from resonance production. Here we take a representative value of $m_V= 1$ eV for the light mediator case so that the low-energy DM-nucleon elastic scattering is not divergent; see also relevant discussions in Sec.~\ref{sec:cons-dm}.} In Sec.~\ref{sec:constraints}, we evaluate the constraints from terrestrial detectors and telescopes mentioned above. We finally conclude in Sec.~\ref{sec:conclusion}.

\section{Elastic and inelastic scatterings between DM and proton}   
\label{sec:xsec}

In this section we study the signatures of collisions between HE CRs and DM inside our Galaxy. Specifically, we consider the elastic scattering that has been commonly studied 
in previous literature  
\begin{equation}
\chi + p \to  \chi + p, 
\label{eq:elas}
\end{equation}
and, more importantly, the inelastic scattering
\begin{equation}
\chi + p \to  \chi + X \to \chi + {\rm hadronic~showers} + \gamma\text{-rays} + {\rm neutrinos}. 
\label{eq:inelas} 
\end{equation} 

Since CRs span a very wide range in energy, 
the energies of the resulting DM as well as the secondary $\gamma$-rays and neutrinos can be very high. 
Therefore, we expect that terrestrial neutrino/DM experiments or $\gamma$-ray telescopes sensitive to different energy regions can all be used to probe/constrain the DM model.

\subsection{Basic formalisms for DM--proton scatterings}
\label{sec:scattering} 

{For convenience, we}
choose to describe DM--proton scattering in the rest frame of the initial proton.
We introduce the four-momenta, $p_\chi=(\widetilde E_\chi, \widetilde{\bf p}_\chi)$, $ p'_\chi=(\widetilde E'_\chi, \widetilde {\bf p}'_\chi)$, $q =  p_\chi-p'_\chi$, $ p_p=(m_N,0)$, and $p_X$ as the incoming momentum of DM, the outgoing momentum of DM, the momentum transfer, the initial momenta of the proton, and the momentum of the outgoing hadronic final state $X$, respectively. 
For the case of elastic scattering, $p_X$ is simply the momentum of the outgoing proton. Note that quantities such as energies and three-momenta with a tilde always refer to the rest frame of the initial proton.         

The differential cross section for elastic scattering is given by \cite{Uehling:1954wp,Cao:2020bwd}
\begin{align}
 \frac{d\sigma_{\rm el}}{dQ^2} = \frac{ g_B^2g_\chi^2}{4\pi \widetilde \beta^2} \frac{1}{(Q^2+m_V^2)^2} \left[ 1 - \widetilde \beta^2 \frac{Q^2}{Q^2_{\rm max}} + \frac{Q^4}{8 m_N^2 \widetilde E_\chi^2} \right] G^2_p(Q^2)\;,  \label{eq:dxsec_elas}  
\end{align}
where $Q^2 \equiv -q^2$ is the positive four-momentum transfer squared, $\widetilde\beta=|\widetilde {\bf p}_\chi|/\widetilde E_\chi$ is the incoming velocity of DM, and $G_p$ is the form factor of proton. 
The maximal momentum transfer squared, $Q^2_{\rm max}$, is 
\begin{align}
 Q^2_{\rm max} = \frac{4(\widetilde \Gamma^2-1)m_\chi^2 m_N^2}{m_\chi^2 + m_N^2 + 2 m_N \widetilde E_\chi}\;, \label{eq:Qmax}
\end{align}
with $\widetilde \Gamma = \widetilde E_\chi/m_\chi=(1-\widetilde \beta^2)^{-1/2}$. As in Refs. \cite{Bringmann:2018cvk,Cappiello:2019qsw}, we assume that $G_p$ takes a dipole form,
\begin{align}
G_p(Q^2) = \frac{1}{\big(1+Q^2/\Lambda_p^2\big)^2}\;,  
\label{eq:form}
\end{align}
with $\Lambda_p \approx$ 770 MeV \cite{Angeli:2004kvy}.

Similarly to neutrino--nucleon scattering \cite{Formaggio:2013kya}, DM--proton inelastic scatterings can be divided into two main categories: resonance excitation of the nucleon (RES) and deep inelastic scattering off the individual quark constituents of the nucleon (DIS). With a lower threshold, resonance production contributes more significantly at $\widetilde E_\chi \lesssim$ a few GeV, and then DIS starts to dominate at higher energies. To describe the inelastic scattering, one can introduce the following quantities (see, e.g., \cite{Zuber}):        
\begin{subequations}
\begin{equation}
\begin{aligned}
\nu = \frac{q \cdot p_p}{m_N} = \widetilde E_\chi - \widetilde E_\chi'\;,
\end{aligned}
\end{equation}
\begin{equation} 
\begin{aligned} 
W^2 = p_X^2 = (p_\chi+p_p-p'_\chi)^2 = m_N^2+2 m_N \nu-Q^2\;,
\label{eq:W}
\end{aligned} 
\end{equation}
\begin{equation}
\begin{aligned}
x = \frac{Q^2}{2q \cdot p_p} = \frac{Q^2}{2 m_N \nu}\;,
\end{aligned}
\end{equation}
\begin{equation}
\begin{aligned}
y = \frac{q\cdot p_p}{p_\chi \cdot p_p}=\frac{\nu}{\widetilde E_\chi}=1-\frac{\widetilde E'_\chi}{\widetilde E_\chi} = \frac{Q^2}{2m_N \widetilde E_\chi x}\;,  
\end{aligned}
\end{equation}
\end{subequations}
where $\nu$ is the energy transfer of DM in the proton rest frame, $W$ is the invariant mass of the hadronic final state $X$, and $x$ and $y$ are the Bjorken variables introduced for DIS. All these four variables are Lorentz invariant quantities.

For the excitation of a spin-$1/2$ resonance state, the hadronic vector current is given by \cite{Leitner:2009zz}
\begin{align}
H^{\mu}_{R,1/2} = \Big\langle R(p_X)\Big| J^{ \mu}_{R,1/2}(Q^2) \Big|p(p_p) \Big\rangle  = \bar u_R(p_X) \Gamma^{\mu}_{R,1/2}(Q^2) u_p(p_p)\;, 
\end{align} 
where $u_p$ and $u_R$ are the Dirac spinors for the proton and the resonance state $R$, respectively, and the vertex function $\Gamma^{ \mu}_{R,1/2}=V^\mu_{R,1/2}$ and $V^\mu_{R,1/2} \gamma^5$ for resonance states with positive and negative parities, respectively. The vector part $V^\mu_{R,1/2}$ can be further parametrized in terms of transition form factors as  
\begin{align}
V^\mu_{R,1/2}(Q^2) = \frac{{\cal F}_1^R(Q^2)}{4 m_N^2} 
\Big(Q^2\gamma^\mu+\slashed{q} q^\mu\Big) + 
\frac{{\cal F}_2^R(Q^2)}{2m_N} i \sigma^{\mu\nu} q_\nu\;.     
\end{align}

The formalism for the excitation of a spin-$3/2$ resonance state (dominated by the $\Delta$-resonance) is more involved. The hadronic current is \cite{Leitner:2009zz} 
\begin{align}
H^\mu_{R,3/2} = \Big\langle R(p_X)\Big| J^{ \mu}_{R,3/2}(Q^2) \Big|p(p_p) \Big\rangle  = \bar \psi_{R,\alpha}(p_X) \Gamma^{\alpha\mu}_{R,3/2}(Q^2) u(p_p)\;,     
\end{align}
with $\psi_{R,\alpha}$ the Rarita-Schwinger spinor for a spin-$3/2$ state. Introducing $\Gamma^{\alpha\mu}_{R,3/2}=V^{\alpha\mu}_{R,3/2}\gamma^5$ and $V^{\alpha\mu}_{R,3/2}$ for positive and negative parities, we have 
\begin{align}
 V^{\alpha\mu}_{R,3/2} =& \frac{C_3^R(Q^2)}{m_N} \Big(g^{\alpha\mu}\slashed q-q^\alpha\gamma^\mu\Big) + \frac{C_4^R(Q^2)}{m_N^2}\Big(g^{\alpha\mu} q \cdot p_X - q^\alpha p_X^\mu\Big) \nonumber \\ 
 & + \frac{C_5^R(Q^2)}{m_N^2}\Big(g^{\alpha\mu} q \cdot p_p - q^\alpha p_p^\mu\Big) + g^{\alpha\mu} C_6^R(Q^2)\;,      
\end{align}
with $C_{3,4,5}^R$ the transition form factors from the proton to the spin-$3/2$ states. 

Once the transition form factors to all relevant resonance states are known, the transition matrix elements and the cross sections for resonance productions can be obtained. Taking into account the propagator of the new vector mediator, the differential cross section can be expressed as 
\begin{align}
 \frac{d^2\sigma_{\rm RES}}{d\nu dQ^2} = \frac{9g_\chi^2 g_q^2}{(Q^2+m_V^2)^2} \frac{ \sum_R |{\cal T}_R|^2 {\cal A}_R}{32\pi m_N |\widetilde{\bf p}_\chi|^2}\;, \label{eq:ddxsec_res} 
\end{align}
where $|{\cal T}_R|^2$ is obtained by contracting the hadronic tensor with that associated with DM, and ${\cal A}_R$ is the spectral function incorporating the widths of the resonance state $R$ (see more details in Ref.~\cite{Leitner:2009zz}). In this work, we simply use the neutrino generator encoded in \texttt{GiBUU}~\cite{Buss:2011mx} to calculate the differential cross section $d^2\sigma_{\rm RES}/d\nu dQ^2$ given in Eq.~\eqref{eq:ddxsec_res}.
Specifically, 
we consider 30 resonance states with masses between 1 and 2 GeV in \texttt{GiBUU} and use the electromagnetic (EM) form factors taken from the MAID analysis \cite{Tiator:2006dq,Drechsel:2007if}. Note that here we have assumed that the vector form factors related to the new vector mediator are well approximated by the EM form factors.

\begin{figure*}[htbp]
\begin{centering}
\subfloat[]{
\includegraphics[width=0.49\textwidth]{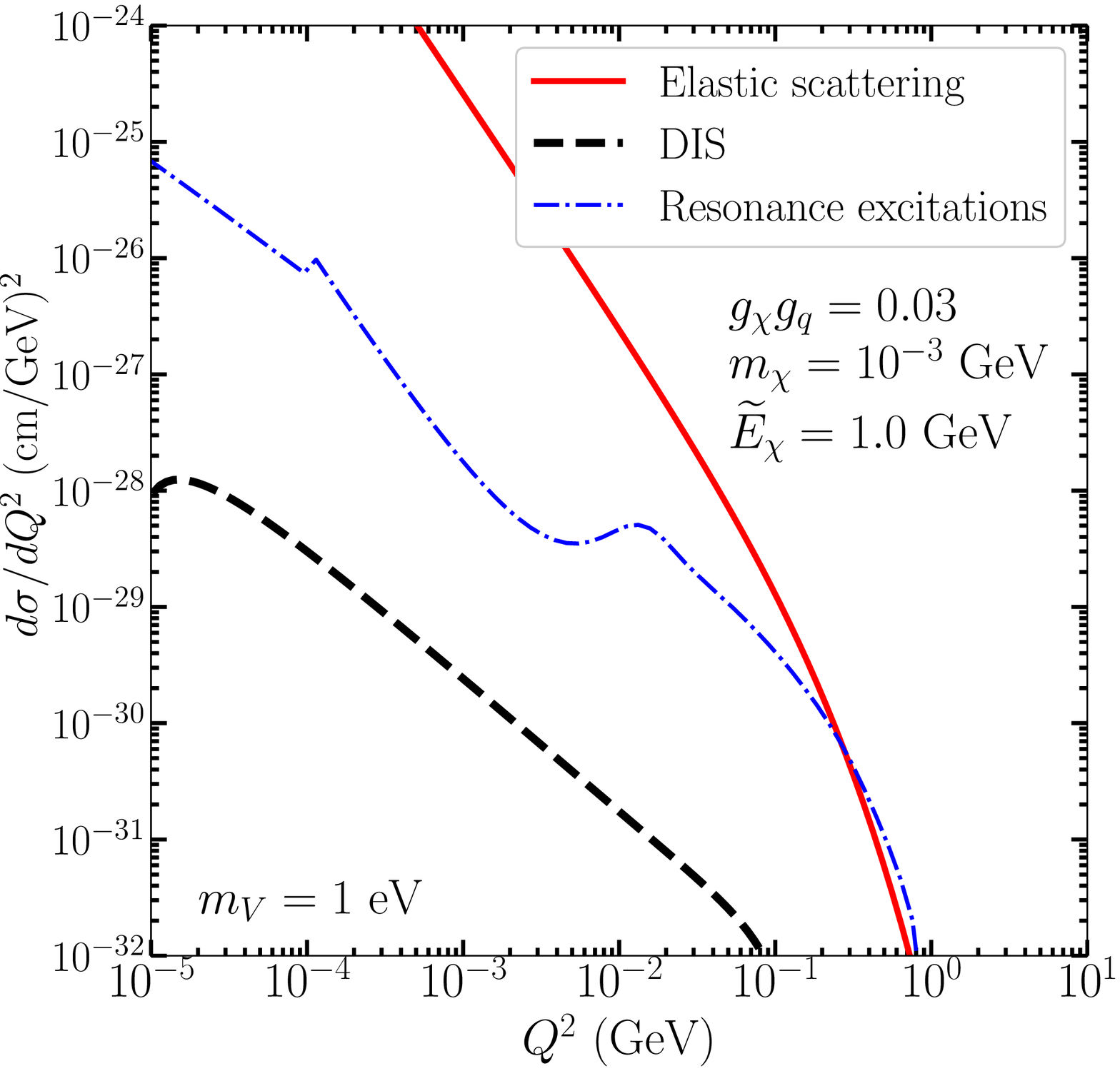}
}
\subfloat[]{
\includegraphics[width=0.49\textwidth]{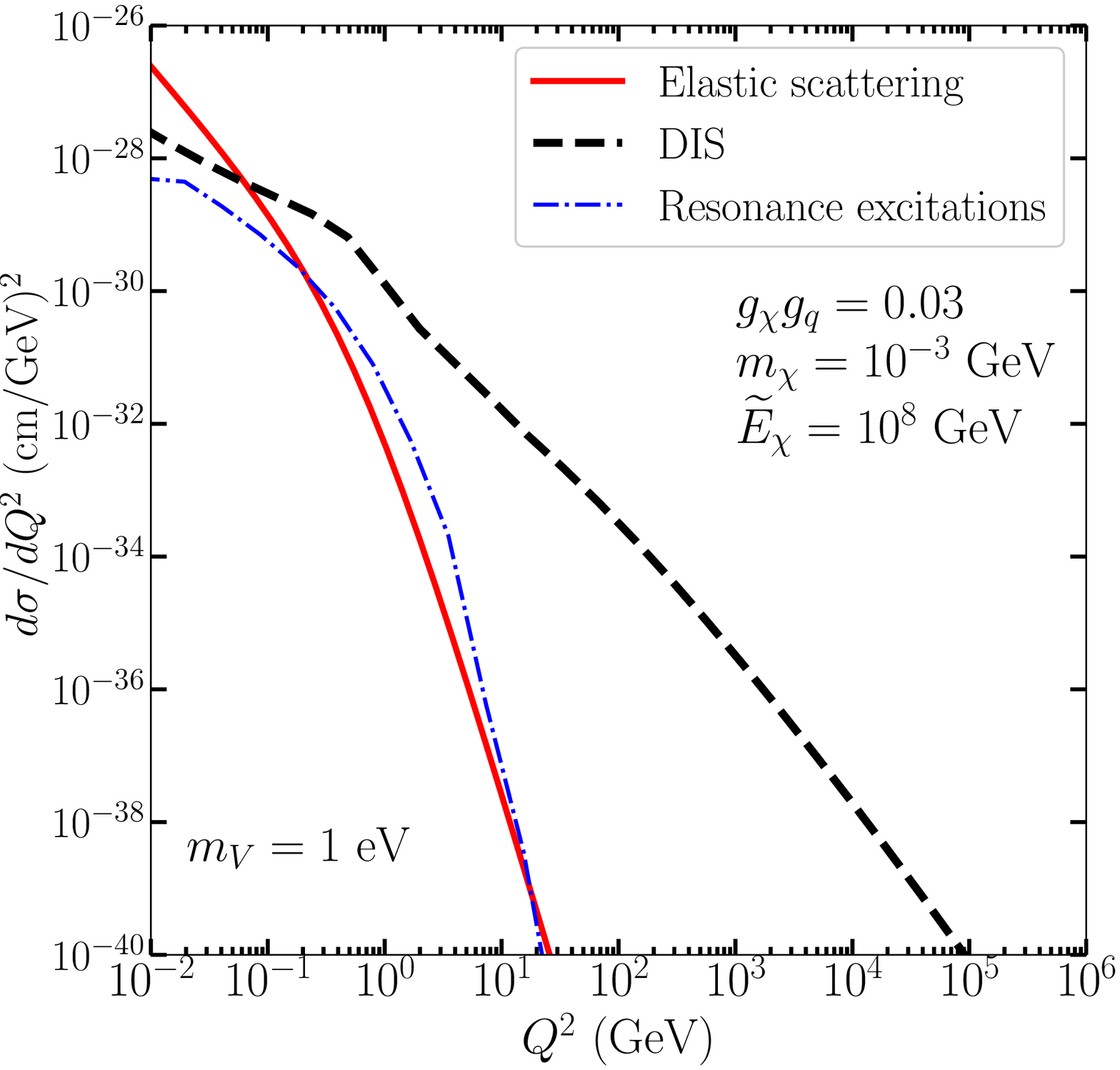}
}
\caption{
The differential cross sections $\dfrac{d\sigma}{d Q^2}$ of elastic scattering, resonance excitation, and DIS as functions of $Q^2$ 
for $\widetilde E_\chi=1\gev$ [panel (a)] and $\widetilde E_\chi=10^{8}\gev$ [panel (b)] with $\widetilde E_\chi$ the DM energy in the proton rest frame. 
%\mkgreen{Because of two very different $\widetilde E_\chi$, we depict the x-axis ($Q^2$) for two panels in their corresponding scales.}
}
\label{fig:DM_dxs}
\end{centering}
\end{figure*}

\begin{figure*}[htbp]
\begin{centering}
% \subfloat[]{
% \includegraphics[width=0.5\textwidth]{massive_Xsec_mx1e-6}
% }
% \subfloat[]{
% \includegraphics[width=0.5\textwidth]{massless_Xsec_mx1e-6}
% }\\
\subfloat[]{
\includegraphics[width=0.5\textwidth]{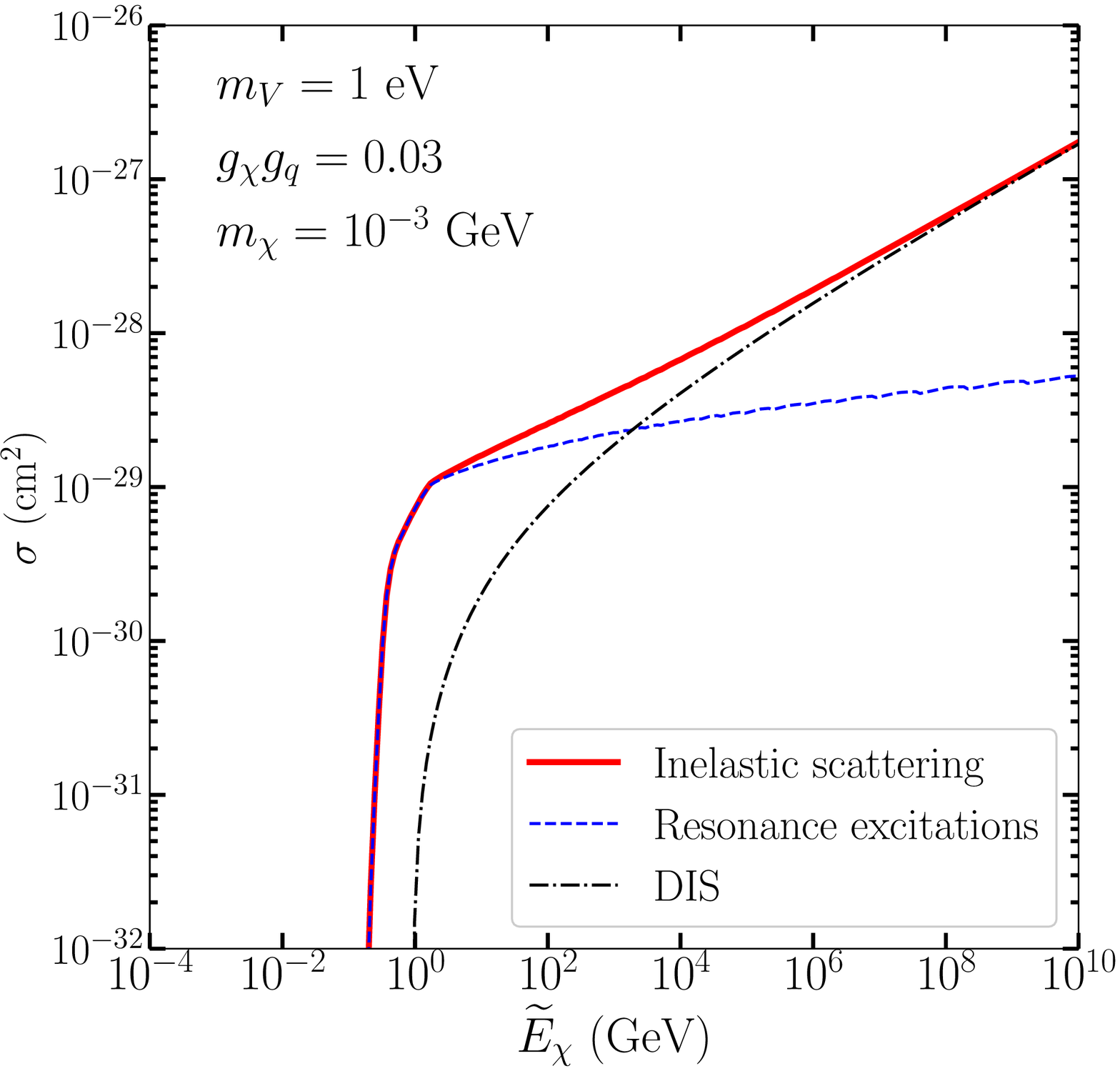}
}
\subfloat[]{
\includegraphics[width=0.5\textwidth]{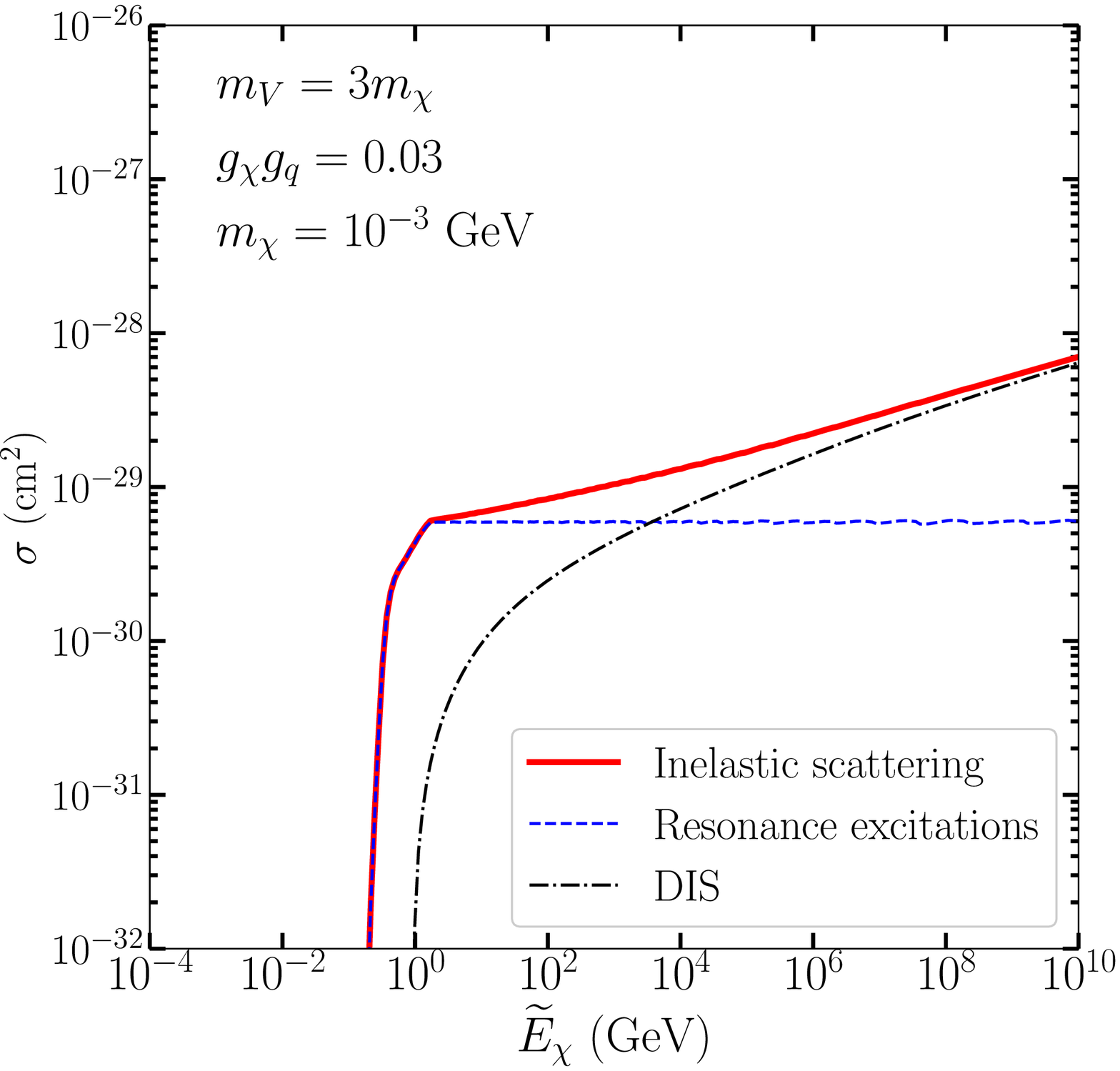}
}
\caption{The total cross sections of inelastic scattering as functions of $\widetilde E_\chi$ 
for the scenarios $m_{V}=1$ eV [panel (a)] and $m_{V}=3 m_\chi$ [panel (b)].
}
\label{fig:DM_xs}
\end{centering}
\end{figure*}

For DIS, the differential cross section is given by 
\begin{align}
\frac{d^2\sigma_{\rm DIS}}{d\nu dQ^2} = \frac{1}{2m_N \widetilde E_\chi^2 y} \frac{d^2\sigma_{\rm DIS}}{dxdy}\;,  
\label{eq:ddxsec_dis}
\end{align}
where 
\begin{align}
\frac{d^2\sigma_{\rm DIS}}{dxdy} = \frac{g_\chi^2 g_q^2}{4\pi(Q^2+m_V^2)^2} \left(1-y+\frac{y^2}{2} - \frac{xy m_N}{2\widetilde E_\chi} \right)F_2(x, Q^2)\;,    
\end{align}
and the corresponding structure function is    
\begin{align}
F_2(x,Q^2) = x \sum_{f=u,d,c,s,b} \Big[f(x,Q^2) + \bar f(x,Q^2)\Big]\;.       
\end{align}
with $f$ $(\bar f)$ the parton distribution functions (PDFs) of quarks (anti-quarks). In our study, we use the MSTW 2008 NNLO PDFs from \cite{MSTW,Martin:2009iq}.

To compare the differential cross sections of the elastic and inelastic scatterings, one can further integrate out the energy transfer $\nu$ in the double differential cross sections [see Eq.~\eqref{eq:ddxsec_dis}] and obtain 
\begin{align}
\frac{d\sigma_{\rm RES,DIS}}{dQ^2} = \int \frac{ d\sigma_{\rm RES,DIS}}{d\nu dQ^2} d\nu\;. \label{eq:dxsec_inel}
\end{align}
Fig. \ref{fig:DM_dxs} shows the differential cross sections $d\sigma/dQ^2$ of 
elastic scattering, resonance excitation, and DIS, respectively for cases with $m_\chi=10^{-3}$ GeV, $g_\chi g_q = 0.03$, and a light vector mediator with $m_V=1$ eV. The differential cross sections for the case with $m_V=3m_\chi$ are very close to the light mediator case, as $Q^2$ explored here is much larger than $m_\chi^2$. For illustration, we have used $\widetilde E_\chi=1$ and $10^8$ GeV as representative low and high energies, respectively. As shown in the figures, all the differential cross sections drop with increasing $Q^2$ at large $Q^2$ mainly due to the factor $Q^{-4}$ from the propagator of the mediator. Besides that, elastic scattering and resonance excitations are further suppressed by a factor of $\sim$$Q^{-8}$ from the hadronic form factors at $Q^2 \gtrsim 1 {\rm GeV}^2$ [see Eq.~\eqref{eq:form}], making DIS more important at higher $Q^2$.
Related to the values of $Q^2$ probed at different $\widetilde{E}_\chi$, $d\sigma_{\rm DIS}/dQ^2$ decreases as $\sim$$Q^{-2}$ at $Q^2 \lesssim 1~{\rm GeV}^2$ and as $\sim$$Q^{-4}$ at higher $Q^2$. As the typical values of $Q^2$ increase with $\widetilde E_\chi$, DIS gives a dominant contribution at high $E_\chi$. This can also be clearly seen in the total cross sections shown in Fig.~\ref{fig:DM_xs}, with $\sigma_{\rm RES,DIS}=\int dQ^2 d\sigma_{\rm RES,DIS}/dQ^2$. With a finite mediator mass $m_V=3 m_\chi$ [see Fig.~\ref{fig:DM_xs}(b)], the total cross section is reduced when compared to that with a light mediator with $m_V=1$ eV [see Fig.~\ref{fig:DM_xs}(a)]. The reductions of the total cross sections of both resonance excitations and DIS are mainly due to the suppressed contributions of $d\sigma_{\rm RES,DIS}/dQ^2$ at $Q^2 \lesssim m_V^2$, with $Q^2_{\rm min}$ the order of $m_\chi^2$. We do not show the total cross section for elastic scatterings as it requires a low-$Q^2$ cutoff to avoid the divergence in the case of $m_V=0$, and is irrelevant to our studies below.

\subsection{DM accelerated by cosmic-rays} 
\label{sec:dm_flx}

DM can be accelerated to high velocities/energies through collisions with HE CRs. In principle, both elastic scatterings and inelastic scatterings contribute to the accelerated DM flux. The scenario involving only the 
elastic scattering has been well studied in previous literature \cite{Bringmann:2018cvk,Ema:2018bih,Cappiello:2019qsw,Guo:2020drq,Ge:2020yuf,Zhang:2020htl,Bondarenko:2019vrb,Dent:2019krz,Wang:2019jtk,Cho:2020mnc,Cao:2020bwd,Jho:2020sku,Bloch:2020uzh}. 
In this subsection we include the contributions from the inelastic channels 
and discuss their possible implications.

The differential DM flux per solid angle $\Omega$ (in units of ${\rm GeV^{-1}~cm^{-2}~s^{-1}~sr^{-1}}$) upscattered by CRs inside the Milky Way and arriving at the Earth is given by a line-of-sight (l.o.s.) integral \cite{Bringmann:2018cvk,Ema:2018bih}
\begin{align}
\phi_\chi^{\rm MW} (T_\chi,\Omega) \equiv \frac{d^2N_\chi}{dT_\chi d\Omega} (T_\chi,\Omega) = \int_{\rm l.o.s.} d\ell \int dE_p \frac{\rho_\chi(r)}{m_\chi}\phi_p(E_p) \left[ \frac{d\sigma_{\rm el}}{dT_\chi} + \frac{d\sigma_{\rm inel}}{dT_\chi} \right],  
%D_{p\chi}(E_p, E_\chi) \sigma_{p\chi}, 
\label{eq:flx_chi} \end{align}
where $T_\chi$ is the kinetic energy of the accelerated DM, $\phi_p \equiv d^2N_p/(dE_p d\Omega)$ is the differential flux of CR protons (in units of ${\rm GeV^{-1}~cm^{-2}~s^{-1}~sr^{-1}}$), and
$\rho_\chi$ is the mass density of the DM halo. Following Refs.~\cite{Ema:2018bih,Cappiello:2019qsw,Guo:2020drq}, we have assumed that the galactic HECRs are uniformly and isotropically distributed in a cylinder with radius $R= 10$ kpc and half-height $h=1$ kpc. This is a relatively good approximation when compared to the CR distribution simulated by \texttt{GALPROP} \cite{Strong:1998pw} and results in only small differences for the accelerated DM flux (see later discussion and Fig.~\ref{fig:relative_flx}).  
To cover a wide energy range of CRs, we adopt the same flux for CR proton as in Ref.~\cite{Guo:2020drq} for our numerical study.
The CR flux can be approximately described by a broken power law, $\phi_p \propto E_p^{-\gamma}$, with $\gamma \approx 2.7$ for $E_p \lesssim 10^6$ GeV (below the knee), $\gamma\approx 3$ for $10^6 \lesssim E_p \lesssim 2\times 10^{8}$ GeV (below the 2nd knee), $\gamma \approx 3.3$ for $2\times 10^8 \lesssim E_p \lesssim 5\times 10^{9}$ GeV (below the ankle), which then followed by a flattening of the CR flux and finally a rapid drop above $3\times 10^{10}$ GeV due to Greisen-Zatsepin-Kuzmin (GZK) cutoff \cite{Zyla:2020zbs}. As we aim to explore the inelastic effects, we neglect the contributions from helium and other heavy isotopes in the CRs. For $\rho_\chi$, we take the Navarro-Frenk-White (NFW) profile \cite{Navarro:1995iw,Navarro:1996gj} with a scale radius $r_s=20$~kpc, 
normalized to the local DM density $\rho_0=0.3\gev~{\rm cm}^{-3}$ where the Sun is located at $r=8.2~{\rm kpc}$ from the galactic center (GC). 
We assume a single-component DM scenario in this work so that the DM local density is entirely made of $\chi$. With DM initially at rest and, thus, $T_\chi = Q^2/(2m_\chi)$ in the laboratory frame, the differential cross sections in Eq.~\eqref{eq:flx_chi} are given by 
\begin{align}
& \frac{d\sigma_{\rm el}}{dT_\chi} = 2 m_\chi \frac{d\sigma_{\rm el}}{dQ^2}\;,  \\ 
& \frac{d\sigma_{\rm inel}}{dT_\chi} = 2m_\chi  \frac{d\sigma_{\rm inel}}{dQ^2} = 2m_\chi \left( \frac{d\sigma_{\rm RES}}{dQ^2} + \frac{d\sigma_{\rm DIS}}{dQ^2}\right)\;,        
\end{align}
where $\frac{d\sigma_{\rm el}}{dQ^2}$ and $\frac{d\sigma_{\rm RES,DIS}}{dQ^2}$ are given in Eqs.~\eqref{eq:dxsec_elas} and \eqref{eq:dxsec_inel}.

\begin{figure*}[htbp]
\begin{centering}
\subfloat[]{
\includegraphics[width=0.51\textwidth]{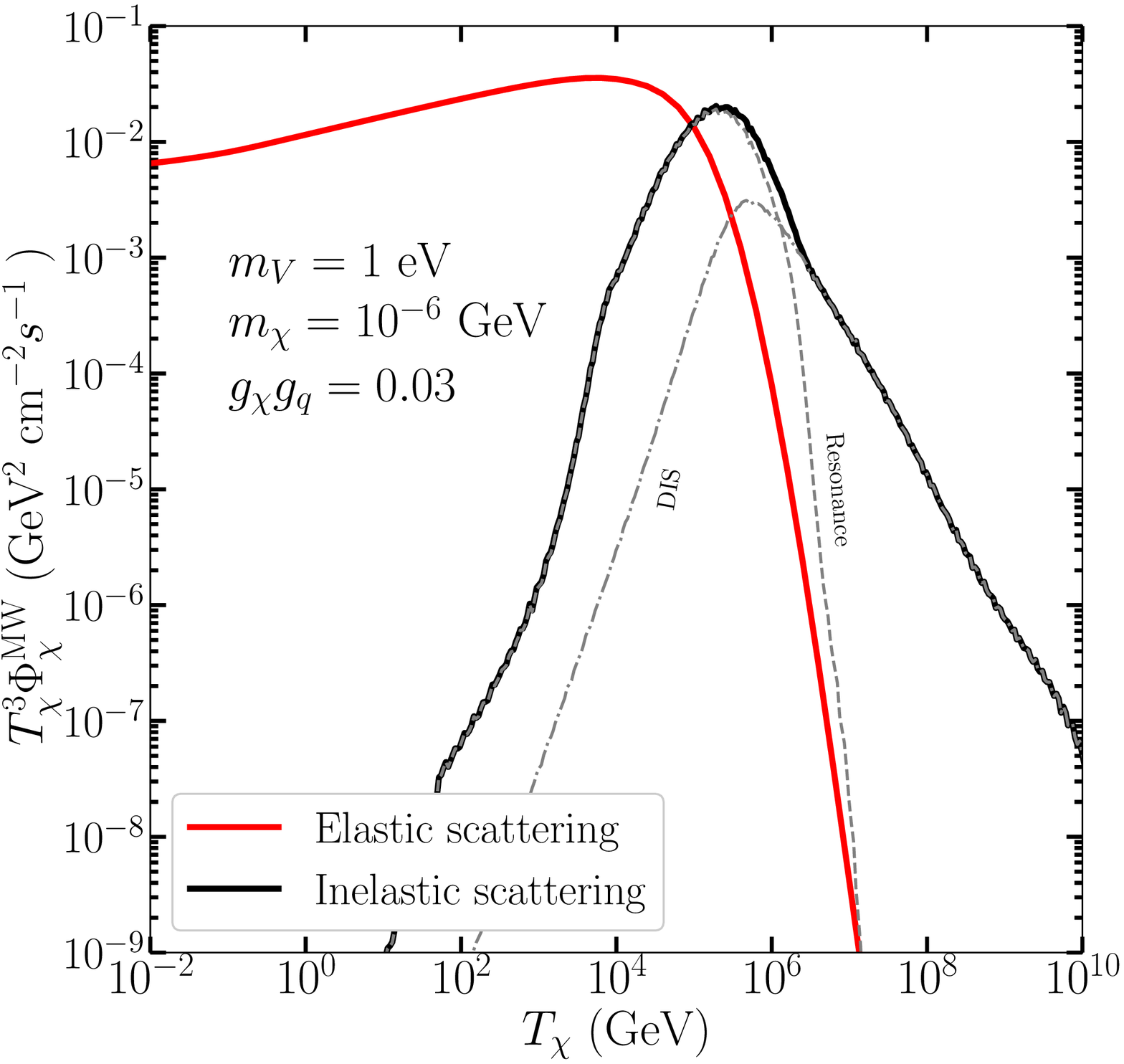}
}
\subfloat[]{
\includegraphics[width=0.51\textwidth]{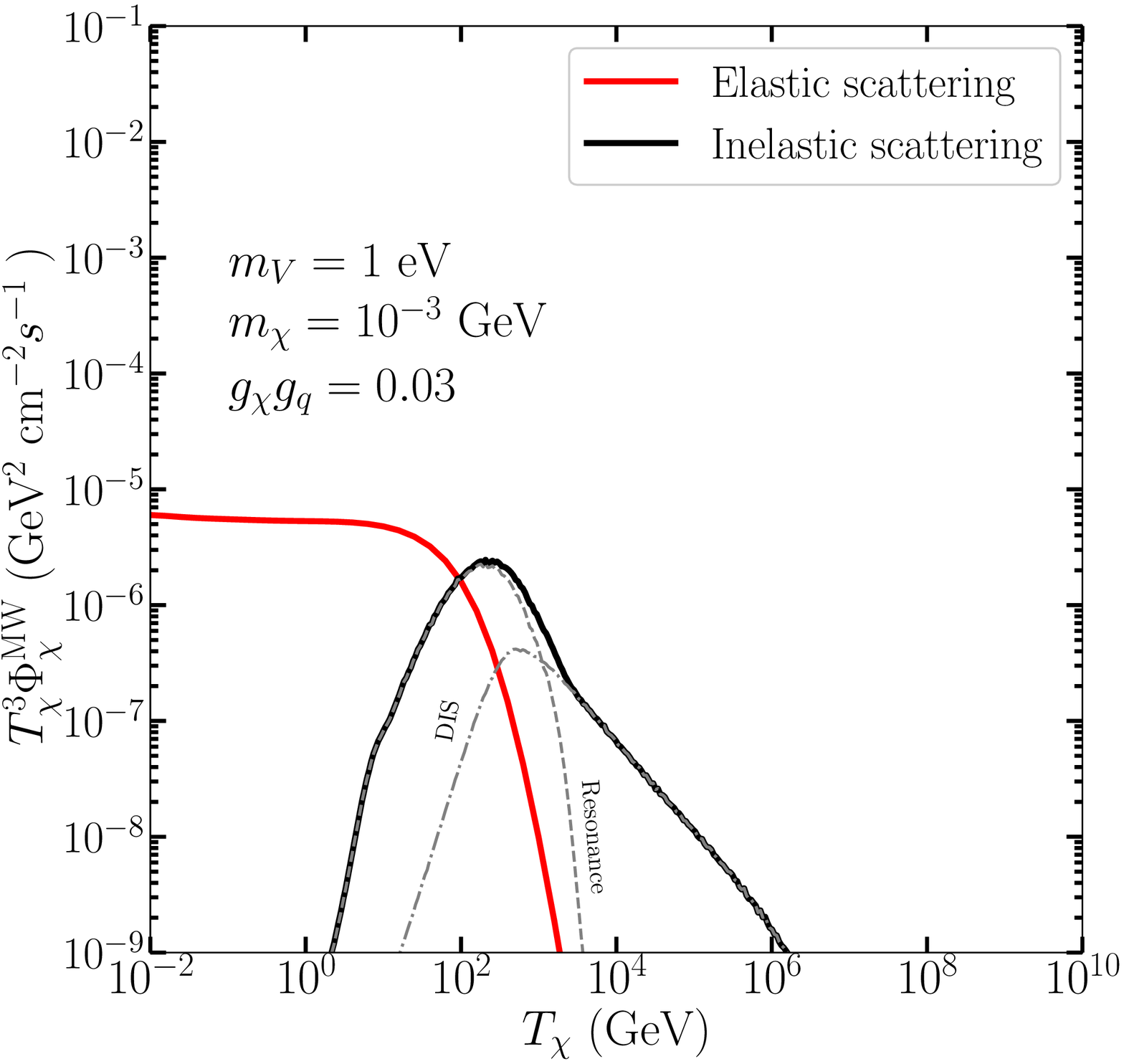}
}
\caption{
The DM fluxes from elastic scattering, resonance productions, and DIS as functions of $T_\chi$ for $m_\chi=10^{-6}\gev$ [panel (a)] and $m_\chi=10^{-3}\gev$ [panel (b)]. 
We take the scenario $m_V=1$ eV as a representative example. The scenario $m_{V}=3 m_\chi$ has a similar distribution. 
}
\label{fig:DM_flx}
\end{centering}
\end{figure*}

Fig.~\ref{fig:DM_flx} shows the boosted DM fluxes from elastic and inelastic scatterings between CRs and DM inside our Galaxy, where we have introduced the solid-angle integrated flux $\Phi_\chi^{\rm MW}$ (in units of ${\rm GeV^{-1}~cm^{-2}~s^{-1}}$) as 
\begin{align}
 \Phi_\chi^{\rm MW}(T_\chi) = \int d\Omega ~\phi_\chi^{\rm MW}(T_\chi, \Omega)\;. 
 \label{eq:phi_DM_4pi}
\end{align}
Note that, to better present the fluxes, we choose to show $T_\chi^3 \Phi_\chi^{\rm MW}$ instead of $\Phi_\chi^{\rm MW}$. We find that at $T_\chi \lesssim 0.1~{\rm GeV}(m_\chi/{\rm GeV})^{-1}$, the fluxes are dominated by elastic scattering. 
For elastic scattering, $d\sigma_{\rm el}/dT_\chi \propto  m_\chi d\sigma_{\rm el}/dQ^2 \propto m_\chi Q^{-4} \propto m_\chi^{-1} T_\chi^{-2}$ at low $Q^2$.
The minimal value of $E_p$ contributing to the flux at $T_\chi$ for $m_\chi < m_N^2/E_p$ is $E_{p,\rm min} \propto (T_\chi/m_\chi)^{1/2}$ [see Eq.~\eqref{eq:Qmax}]. 
Combining these with a CR proton spectrum of $E_p^{-2.7}$, Eq.~\eqref{eq:flx_chi} gives rise to a DM flux at low $T_\chi$ being proportional to $m_\chi^{-1}(E_{p,\rm min})^{-1.7} d\sigma_{\rm el}/dT_\chi \propto m_\chi^{-1.15} T_\chi^{-2.85}$.
At higher $T_\chi$ or $Q^2$, $d\sigma_{\rm el}/dQ^2 \propto Q^{-12}\propto T_\chi^{-6}$. The
corresponding DM fluxes from elastic scattering decrease as $\sim$$T_\chi^{-6.85}$ as shown in Fig.~\ref{fig:DM_flx}. At $T_\chi \gtrsim 0.1~{\rm GeV}~(m_\chi/{\rm GeV})^{-1}$, resonance excitations and DIS start to contribute dominantly. Based on similar arguments, the fluxes above $\sim$1~${\rm GeV}~(m_\chi/{\rm GeV})^{-1}$ vary as $m_\chi^{-5.15} T_\chi^{-6.85}$ from resonance excitations similar to that from elastic scattering, and as $m_\chi^{-2} T_\chi^{-3.7}$ from DIS considering that $E_{p, \rm min}$ roughly scales as $T_\chi$.

\begin{figure*}[htbp]
\begin{centering}
\includegraphics[width=0.6\textwidth]{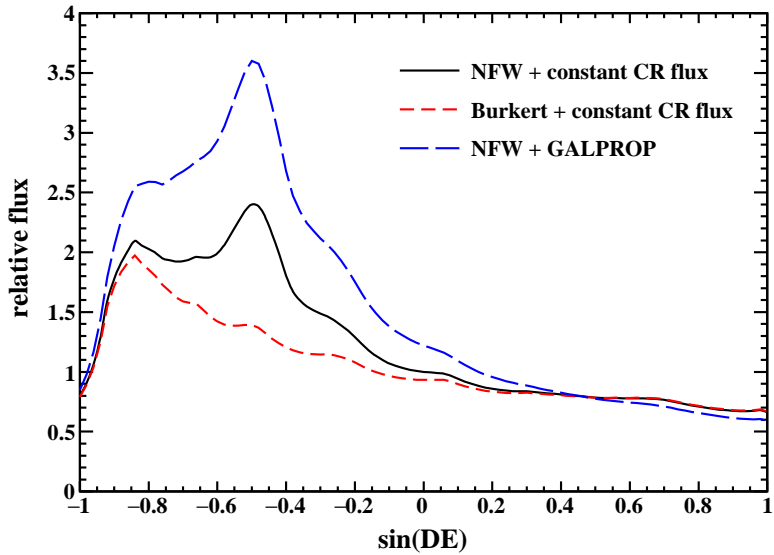}
\caption{Relative angular distribution of DM flux as a function of $\sin(\texttt{DE})$. The flux for the case ``NFW$+$constant CR flux" is normalized to 1 at $\sin(\texttt{DE})=0$. For comparison, the same scaling or normalization factor is taken for the other two cases.
%\mkgreen{The normalization factor in this plot is the flux $\phi^{\rm MW}_{\chi, \texttt{DE}}$ of the case ``NFW$+$constant CR flux" at $\texttt{DE}=0$.}
%\sming{What is the value of $T_\chi$ here?}
}
\label{fig:relative_flx} 
\end{centering}  
\end{figure*}

The upscattered DM flux exhibits a strong angular dependence on both the right ascension (\texttt{RA}) and the declination (\texttt{DE}) in the equatorial coordinate system~\cite{Guo:2020drq,Ge:2020yuf}.
For detectors not located at the North or South poles, this can lead to daily modulations due to its dependence on \texttt{RA}~\cite{Ge:2020yuf}. 
However, when considering a long exposure time, one can integrate over the \texttt{RA}-dependence as done in Ref.~\cite{Guo:2020drq}. Fig.~\ref{fig:relative_flx} shows the relative DM flux integrated over \texttt{RA}, $\phi^{\rm MW}_{\chi,\texttt{DE}} = \int_0^{2\pi} d\texttt{RA} ~\phi_\chi^{\rm MW}(T_\chi, \Omega)$, as a function of $\sin({\texttt{DE}})$ for our fiducial scenario adopting uniformly distributed CRs and the NFW halo profile.
%The full angular distribution in both \texttt{RA} and \texttt{DE} of the CR-accelerated DM fluxes can be found in Ref.~\cite{Guo:2020drq}. 
Note that for this case, we have normalized $\phi^{\rm MW}_{\chi, \texttt{DE}}(T_\chi, \texttt{DE})$ to unity at $\sin(\texttt{DE})=0$. 
Since the angular dependence is determined by the spatial distribution of the DM halo and the location of the Earth, the relative angular distribution is independent of $T_\chi$. 
Also shown in Fig.~\ref{fig:relative_flx} is the relative flux using the Burkert halo profile \cite{Burkert:1995yz} with the uniform CR flux. Note that we take the same normalization factor as for the fiducial case. For the Burkert profile, we use a scale radius of 9 kpc and the same local density as the NFW profile. In reality, the CR flux
varies at different positions inside the Milky Way. 
To estimate the associated effects, we assume that the spatial distribution of the CR proton flux is independent of the CR energy. We then use the distribution at $E_p=1$ TeV obtained from \texttt{GALPROP} (see Ref.~\cite{Ge:2020yuf}) for all values of $E_p$, and calculate the relative DM flux with the spatial-dependent CR flux and the NFW profile (see the blue dashed line). Obviously, the resulting angular distribution of the DM flux is also independent of $T_\chi$ as for the case with a constant CR flux. A comparison of these curves in Fig.~\ref{fig:relative_flx} shows that the uncertainty due to the chosen CR model and/or the halo profile is within a factor of $\sim$2. Hereafter, we stick to the fiducial case, i.e, assuming a constant CR flux and the NFW halo profile.

The same angular distribution shown in Fig.~\ref{fig:relative_flx} also applies to the secondary $\gamma$-rays and neutrinos from the inelastic scatterings. 
Although the flux peaks at the direction of the GC for cases with the NFW profile and at the direction of $\sin{\texttt{DE}}\approx -0.84$ with the Burkert profile,
the relative differences at different $\sin(\texttt{DE})$ are
within a factor of $\sim$3. 
Hereafter, we approximate the flux of the upscattered DM and those of the secondary neutrinos as isotropic for the sake of simplicity. With this approximation, we assume that the total flux $\Phi_\chi^{\rm MW}$ in Eq.~\eqref{eq:phi_DM_4pi} is uniformly distributed in all directions \cite{Bringmann:2018cvk,Ema:2018bih}. 
This, on the one hand, simplifies the derivation of 
constraints from the DM flux. 
On the other hand, it allows us to
directly use the existing upper limits on astrophysical or cosmogenic HE neutrinos, which are always assumed to be isotropic.
Considering the full angular dependence could, in principle, enhance the sensitivity of our study.

\subsection{Secondary $\gamma$-rays and neutrinos}  
\label{sec:secondaries}

The fluxes of the secondary $\gamma$-rays and neutrinos summing over all flavors in units of ${\rm GeV^{-1}~cm^{-2}~s^{-1}~sr^{-1}}$ can be expressed as 
\begin{align}
 \phi_{\gamma,\nu}(E_{\gamma,\nu}, \Omega)= 
\int_{\rm l.o.s.} d\ell \int dE_{p} \frac{\rho_\chi(r)}{m_\chi}\phi_p(E_p) 
\int d\nu dQ^2 \times \frac{d^2\sigma_{\rm inel}}{d\nu dQ^2} 
\times \frac{d\mathcal{N}_{\gamma,\nu}(E_p, E_{\gamma,\nu}, \nu, Q^2)}{dE_{\gamma,\nu}}\;,   
\label{eq:gammanu_flx}
\end{align}
where $d\mathcal{N}_{\gamma,\nu}/dE_{\gamma,\nu}$ is the spectrum of produced $\gamma$-rays or neutrinos per $p\chi$ collision for given values of $E_p$, $\nu$, and $Q^2$. We first use \texttt{GIBUU} \cite{Buss:2011mx} and \texttt{SOPHIA} \cite{Mucke:1999yb} to generate the $\gamma$-ray or neutrino yields in the rest frame of the hadronic final state $X$, 
which depend only on the invariant mass $W$ [see Eq.~\eqref{eq:W}] and are isotropically distributed in three-momentum. We then boost the yields from the rest frame of $X$ to the laboratory frame to obtain $d\mathcal{N}_{\gamma,\nu}/dE_{\gamma,\nu}$. The corresponding boost factor is simply $\Gamma_{\rm boost} = E_X/W$ with $E_X=E_p-T_\chi=E_p-Q^2/(2m_\chi)$ the energy of $X$ in the laboratory frame.

To constrain the DM model using the secondary $\gamma$-rays, we consider the observed data from Fermi \cite{TheFermi-LAT:2017vmf} and H.E.S.S. \cite{Abramowski:2016mir} with $E_\gamma$ ranging from $\sim$GeV to $\sim$100~TeV. Both the RES and DIS contribute in 
this energy range. We first rely on \texttt{GIBUU} to generate the $\pi^0$ yields from resonance excitations, and then calculate the resulting $\gamma$-ray yields from $\pi^0 \to \gamma + \gamma$. Note that the $\gamma$-rays in the rest frame of $\pi^0$ carry an energy of $m_\pi/2$ and are isotropically distributed in three-momentum. The resulting $\gamma$-ray yields from $\pi^0$ decay can be obtained by a Lorentz boost, leading to \cite{Hooper:2018bfw}
\begin{align}
\frac{dn_\gamma(E_\gamma)}{dE_\gamma} = \int_{E_{\pi,{\rm min}}(E_\gamma)} dE_\pi \frac{dn_\pi(E_\pi)}{dE_\pi}\frac{1}{\sqrt{E_\pi^2-m_\pi^2}}\;,
\label{eq:pi_decay}
\end{align}
where $E_{\pi,{\rm min}}(E_\gamma)$ satisfies the equation $E_{\pi,{\rm min}}(1+\beta_{\pi,{\rm min}})/2=E_\gamma$ with $\beta_{\pi,{\rm min}} = \sqrt{E_{\pi,{\rm min}}^2-m_\pi^2}\Big/E_{\pi,{\rm min}}$ for $E_\gamma > m_\pi/2$. For relativistic $\pi^0$, $E_{\pi,{\rm min}}(E_\gamma) \approx E_\gamma$, and the resulting $\gamma$-ray takes a same spectral index as the $\pi^0$ flux.
For HE $\gamma$-ray production in the DIS region that may go beyond the scope of \texttt{GIBUU}, we use the \texttt{SOPHIA} code to obtain the $\gamma$-ray yields directly. On the other hand, we consider HE neutrinos above $\sim$PeV that can be probed by IceCube. Since neutrino production involves three-body decays or decay sequences, we directly use the \texttt{SOPHIA} code to calculate the total neutrino yields of all flavors from both resonance production and DIS. We further assume that neutrino fluxes are equally distributed in all flavors after oscillation.   

The $\gamma$-ray fluxes in Eq.~\eqref{eq:gammanu_flx} are further integrated over certain solid angles around the GC to be compared with the observed data at Fermi and H.E.S.S. (see  Fig.~\ref{fig:gamma_flux}). For the secondary neutrinos, we take the isotopic approximation as mentioned in Sec.~\ref{sec:dm_flx} and use the averaged neutrino flux per solid angle to derive the bounds (see Fig.~\ref{fig:nu_flux}).

\section{Constraints from terrestrial experiments and telescope} \label{sec:constraints}

In this section, we derive constraints on the coupling constants of the vector portal model for sub-GeV DM based on the HE $\gamma$-ray observation (Sec.~\ref{sec:cons-gamma}),
the HE neutrino detection (Sec.~\ref{sec:cons-nu}), and 
the expected DM signals at the low-energy DM/neutrino experiments
%capable to probe directly the DM--nucleon interaction
(Sec.~\ref{sec:cons-dm}).

\subsection{Constraints from Fermi and H.E.S.S.} \label{sec:cons-gamma}

To constrain the coupling constants of the vector portal model with the secondary $\gamma$-ray emission, we consider two datasets from Fermi and H.E.S.S..
For the Fermi data, 
we consider the Fermi GC excess at $0.5\gev<E_\gamma<500\gev$~\cite{TheFermi-LAT:2017vmf}  extracted from a circular region within 10$^\circ$ from the GC excluding the inner region of radius $2^\circ$ \cite{TheFermi-LAT:2017vmf}. 
For the HE H.E.S.S. data, we take those with $0.2\tev<E_\gamma<60\tev$, integrated over an annulus centered at Sgr A$^*$ with inner and outer radii of 0.15$^\circ$ and 0.45$^\circ$, respectively, and a section of 66$^\circ$ excluded \cite{Abramowski:2016mir}.

In order to compare the produced secondary $\gamma$-ray fluxes from the inelastic collisions between the CRs and DM, we introduce the 
$J$-factors for Fermi and H.E.S.S. as   
\begin{equation}
  J_{\rm Fermi,~H.E.S.S.} = \int_{\Delta\Omega_{\rm Fermi,~H.E.S.S.}} d\Omega \int_{{\rm l.o.s.}} d\ell~\frac{\rho_{\chi}(r)}{m_\chi}
\end{equation}
to account for different sky coverages in the two data sets mentioned above. The corresponding $J$-factors for Fermi and H.E.S.S. are $J_{\rm Fermi} \approx 2.16~{\rm kpc~cm^{-3}}$ and $J_{\rm H.E.S.S.} \approx 7.47\times 10^{-3}~{\rm kpc~cm^{-3}}$, respectively.
Using these $J$-factors, the expected secondary $\gamma$-ray fluxes in units ${\rm GeV^{-1}~cm^{-2}~s^{-1}}$ are  
\begin{align}
  \Phi_{\rm Fermi,~H.E.S.S.}(E_\gamma) &= \int_{\Delta \Omega_{\rm Fermi,~H.E.S.S.}}d\Omega ~\phi_\gamma(E_\gamma, \Omega)   
  \nonumber \\
  &=J_{\rm Fermi,~H.E.S.S.} \times \int dE_{p} \phi_p(E_p) 
\int d\nu dQ^2 \times \frac{d^2\sigma_{\rm inel}}{d\nu dQ^2} 
\times \frac{d\mathcal{N}_{\gamma}(E_p, E_{\gamma}, \nu, Q^2)}{dE_{\gamma}}\;.  
\end{align}
We point out that these $J$-factors can be taken out, because we
assume that the CR proton flux is uniformly distributed in the Galaxy. 
In reality, the CR flux near the GC could be a factor of a few higher than the local flux, leading to $J$-factors larger than those based on a constant flux by a factor of $\lesssim$2 (see also Fig.~\ref{fig:relative_flx}). 
Therefore, the limits obtained here with $\gamma$-rays can be considered as conservative ones.

\begin{figure*}[htbp]
\begin{centering}
\subfloat[]{
\includegraphics[width=0.51\textwidth]{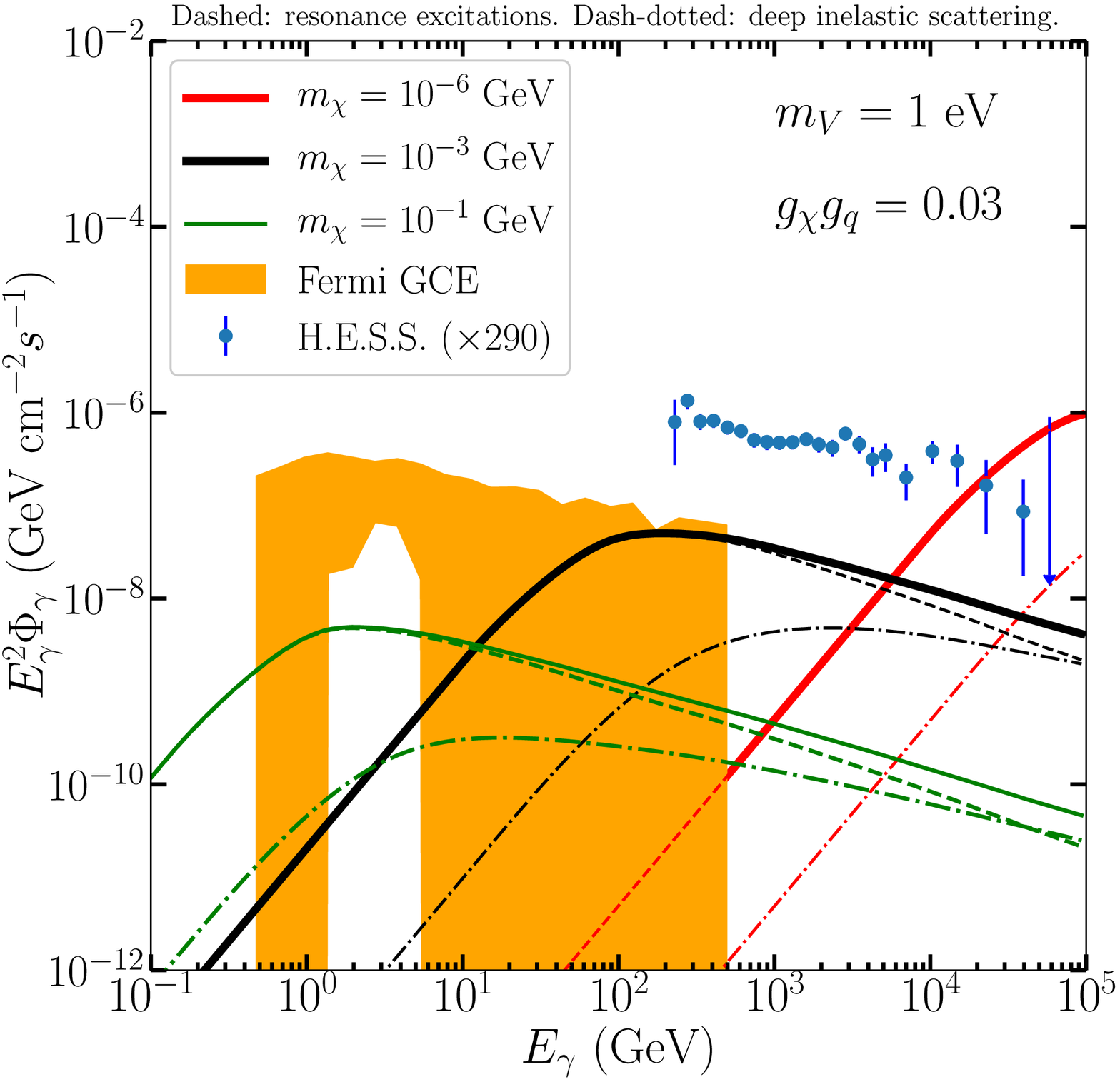}
}
\subfloat[]{
\includegraphics[width=0.51\textwidth]{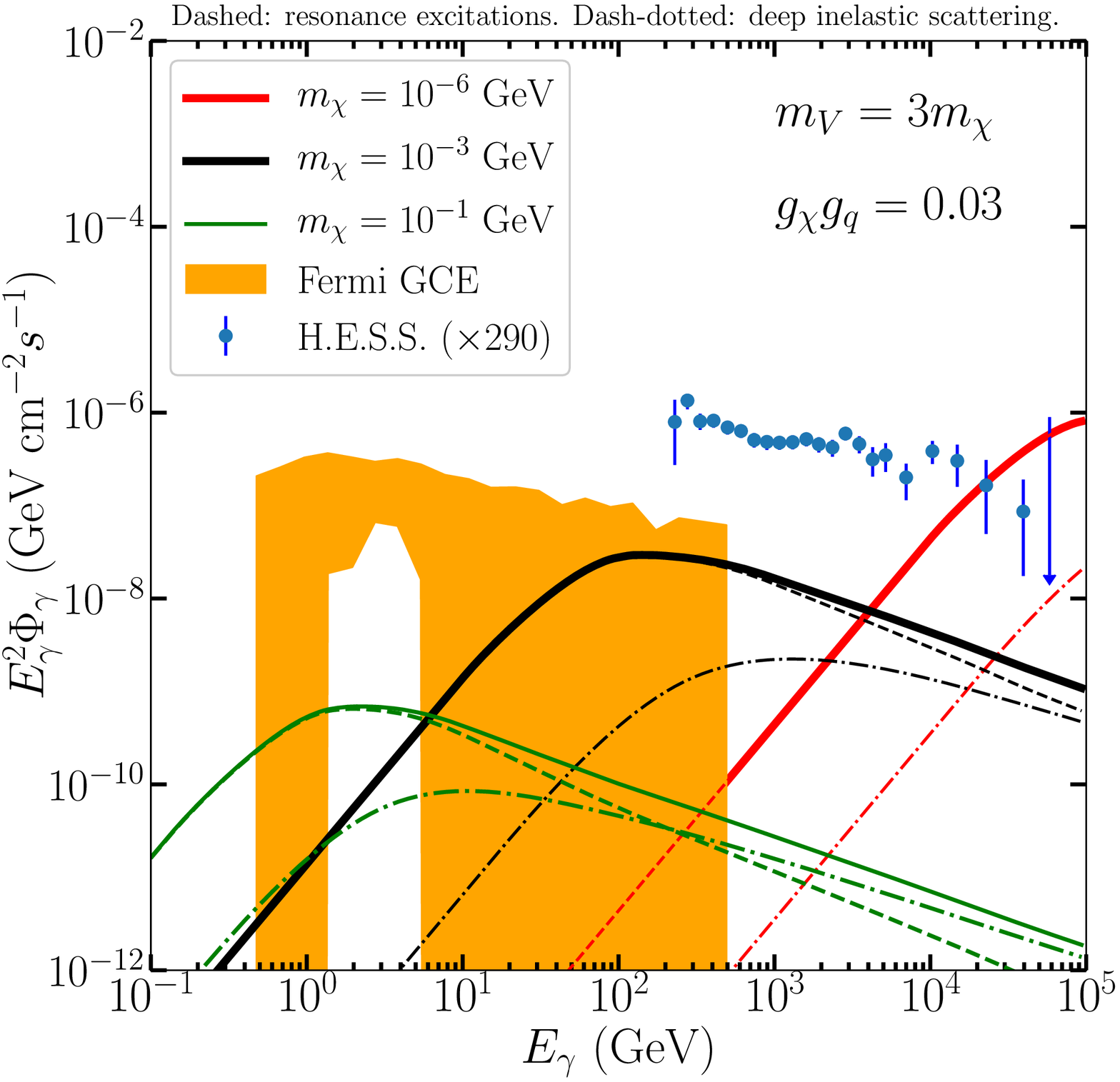}
}
\caption{
The $\gamma$-ray fluxes resulting from $\chi$--$p$ inelastic scattering
for $m_V=1$ eV [panel (a)] and $m_V=3 m_\chi$ [panel (b)]. 
The total inelastic fluxes are presented as solid lines, while 
the dashed and dash-dotted lines are contributions from resonance excitations and 
DIS, respectively. 
}
\label{fig:gamma_flux}
\end{centering}
\end{figure*}

We show the resulting $\gamma$-ray fluxes $\Phi_\gamma(E_\gamma)\equiv\Phi_{\rm Fermi}(E_\gamma)$ in Fig.~\ref{fig:gamma_flux} for different DM masses with $m_V=0$ [panel (a)] and $m_V=3 m_\chi$ [panel (b)]. 
For comparison, the allowed band of the Fermi GC excess taking into account all possible systematic uncertainties is 
shown in Fig.~\ref{fig:gamma_flux} (see Fig.~15 of Ref.~\cite{TheFermi-LAT:2017vmf}).
The flux $\Phi_{\rm H.E.S.S.}(E_\gamma)$ to be confronted with the H.E.S.S. data can be obtained from $\Phi_{\rm Fermi}(E_\gamma)$ by simply multiplying a scaling factor $J_{\rm H.E.S.S.}/J_{\rm Fermi} \approx 3.4\times 10^{-3}$.
Thus, we show instead the rescaled
H.E.S.S. data at $100~{\rm GeV} \lesssim E_\gamma \lesssim 100$ TeV (see Fig.~3 of Ref.~\cite{Abramowski:2016mir}), multiplied by a factor of $J_{\rm Fermi}/J_{\rm H.E.S.S.} \approx 290$ so that they can be directly compared with the calculated $\Phi_{\rm Fermi}$. 

For the range of $E_\gamma$ considered, resonance productions contribute more significantly than DIS. 
As the produced $\gamma$-ray typically carries a certain fraction of the energy of the primary CR proton, the flux given in Eq.~\eqref{eq:gammanu_flx} roughly scales as $\phi(E_\gamma) \propto \phi_p(E_p) \sigma_{\rm RES}(\widetilde{E}_\chi)$ with $\widetilde E_\chi = E_p m_\chi/m_N$; see also discussions below Eq.~\eqref{eq:pi_decay}. Considering that $\sigma_{\rm RES}$ firstly increases rapidly above the threshold and then saturates at $\widetilde E^{\rm sat}_\chi \approx 1$ GeV (see Fig.~\ref{fig:DM_xs}),
the $\gamma$-ray spectra above the peak shown in Fig.~\ref{fig:gamma_flux}, can be understood quantitatively. The flux $\Phi_\gamma(E_\gamma)$ first increases rapidly and then decreases as $E_\gamma^{-2.7}$ above $E^{\rm peak}_\gamma \approx 0.1~{\rm GeV} (m_\chi/{\rm GeV})^{-1}$.
Note that $E^{\rm peak}_\gamma$ is simply determined by $\widetilde E^{\rm sat}_\chi$, as the resulting $E_\gamma$ from $\pi^0$ decay, on average, is $\langle E_\gamma \rangle= 0.5 \langle E_\pi \rangle \approx 0.1E_p = 0.1 \widetilde E_\chi m_N/m_\chi$, i.e., $E_\gamma^{\rm peak} \approx 0.1 \widetilde E^{\rm sat}_\chi m_N/m_\chi \approx 0.1~{\rm GeV}(m_\chi/{\rm GeV})^{-1}$. The simple argument does not apply to the spectra below the peak, where the $\gamma$-ray flux decreases more slowly with decreasing $E_\gamma$ than that would be indicated from the sharply declined $\phi_p \sigma_{\rm RES}$ (see Fig.~\ref{fig:DM_xs}).
The reason is that the decay of $\pi^0$ with energy $E_\pi$ generates a flat distribution of $\gamma$-rays with $E_\gamma^{\rm max,min} = E_\pi(1\pm \beta_\pi)/2$, where $\beta_\pi$ is the velocity of $\pi^0$ in units of the speed of light. Therefore, the $\gamma$-ray flux below the peak receives contributions dominantly 
from $\pi^0$ with $E_\pi\gtrsim E_\pi^0\approx 2E_\gamma^{\rm peak}$ produced with large cross sections.
To understand the $\gamma$-ray yields below the peak, one needs to fully consider the energy spread of $\pi^0$-decay using Eq.~\eqref{eq:pi_decay}. In this case, the lower bound of Eq.~\eqref{eq:pi_decay} is $\sim$$E_\pi^0$ which is independent of $E_\gamma$ and is larger than $E_{\pi,\rm min}(E_\gamma)$. With the $\pi^0$ flux $\propto E_\pi^{-2.7}$ above $E_\pi^0$, the resulting $\gamma$-ray flux below the peak would be independent of $E_\gamma$, or equivalently, $E_\gamma^2\Phi_\gamma \propto E_\gamma^2$, as clearly shown in Fig.~\ref{fig:gamma_flux}.
As $m_\chi$ increases, the peak position of $E_\gamma^2\Phi_\gamma$ shifts to low values of $E_\gamma$ as $E^{\rm peak}_\gamma\propto m_\chi^{-1}$.
Correspondingly, the peak magnitude of $E_\gamma^2\Phi_\gamma$ scales approximately as $m_\chi^{-0.7}$. To derive the upper limits at the 90\% confidence level (C.L.) from the Fermi data, we require that, at any given energy, $\Phi_{\rm Fermi}(=\Phi_\gamma)$ is
smaller than the upper edge of the GC excess band 
multiplied by a factor of 1.28.\footnote{Note that to obtain the 90\% C.L. upper limits, we have simply assumed that the Fermi excess data follow a Gaussian distribution with a mean value of zero and a variance given by the upper edge of the band. For the H.E.S.S. data, we also assume a Gaussian distribution. The upper limits derived in our work are insensitive to these assumptions.}
Similarly, for the H.E.S.S. data, we exclude the regions of large coupling constants if the predicted flux $\Phi_{\rm H.E.S.S.}(\approx \Phi_\gamma/290)$ exceeds the mean values of the measured diffuse flux plus $1.28$ times the deviations at any energy. 
The resulting 90\% C.L. upper limits on $g_\chi g_q$ from the Fermi and the H.E.S.S. data as functions of
the DM mass
are presented in Fig.~\ref{fig:limits}. As $m_\chi$ increases, the peak of $E_\gamma^2 \Phi_\gamma$ moves closer to the energies probed by the $\gamma$-ray telescopes, resulting in tighter bounds.
For $m_\chi\gtrsim 2\times 10^{-6}$ and $\gtrsim 2 \times 10^{-4}$~GeV, the derived limits using the Fermi and the H.E.S.S. data become weaker with $m_\chi$, due to the suppressed $\gamma$-ray fluxes.

The constraints using the $\gamma$-ray fluxes are roughly consistent with the earlier work \cite{Cyburt:2002uw},\footnote{Ref.~\cite{Hooper:2018bfw} focused on dark matter heavier than 1 GeV.} although a constant inelastic scattering cross section was assumed. The reason is that the relevant energy window for $E_\gamma$ is relatively narrow, and besides, the $U(1)_B$ model adopted in our work exhibits a relatively flat inelastic cross section above the threshold. To set a scale, both our studies give an upper bound on the inelastic cross section of a few
 times $10^{-29}~{\rm cm}^2$ for $m_\chi$ around 0.1~GeV using the GC $\gamma$-ray data. Ref.~\cite{Cyburt:2002uw} also checked that using the $\gamma$-ray data off the GC would only give rise to a relaxed bound.

\begin{figure*}[htbp]
\begin{centering}
\subfloat[]{
\includegraphics[width=0.51\textwidth]{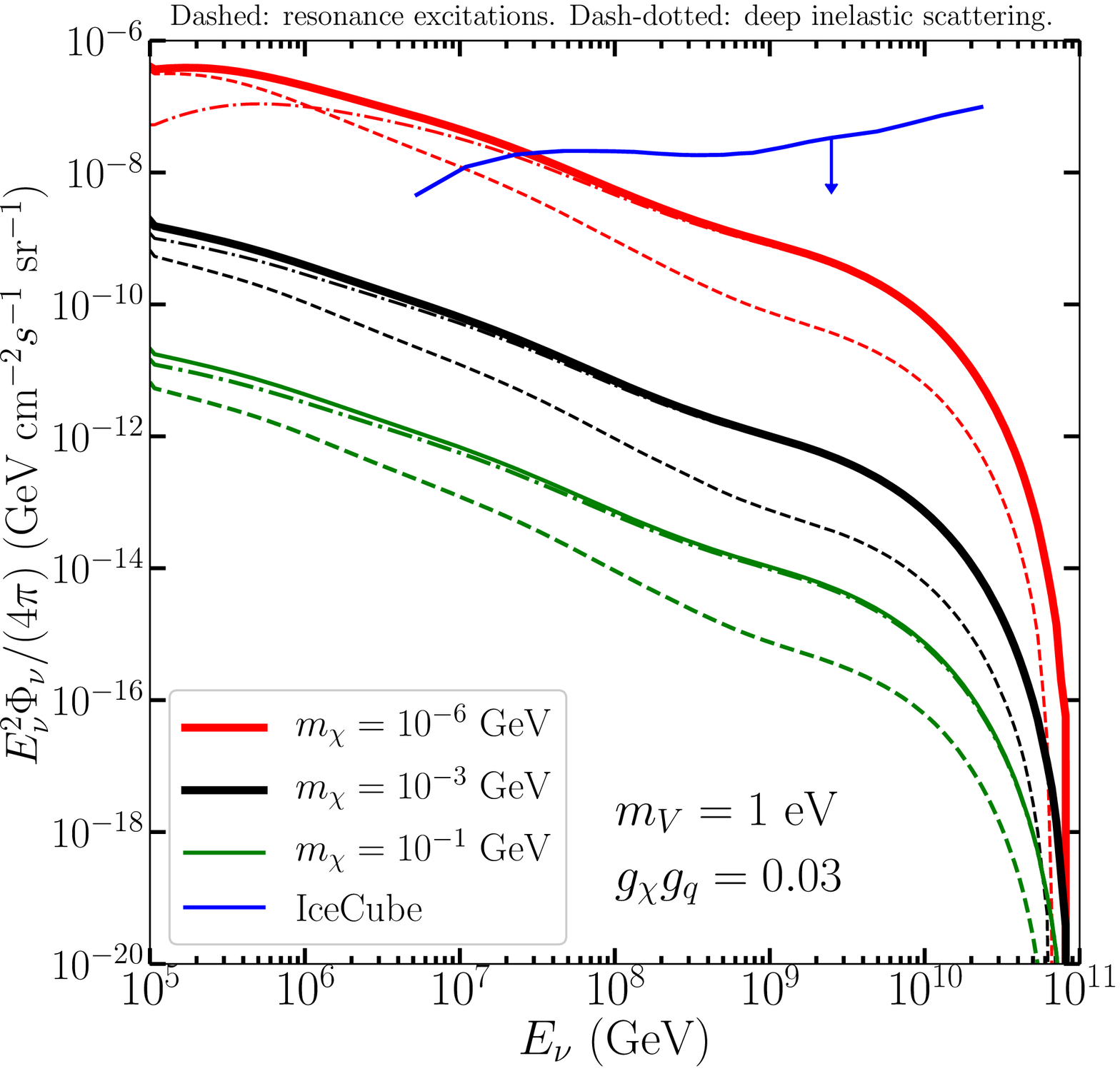}
}
\subfloat[]{
\includegraphics[width=0.51\textwidth]{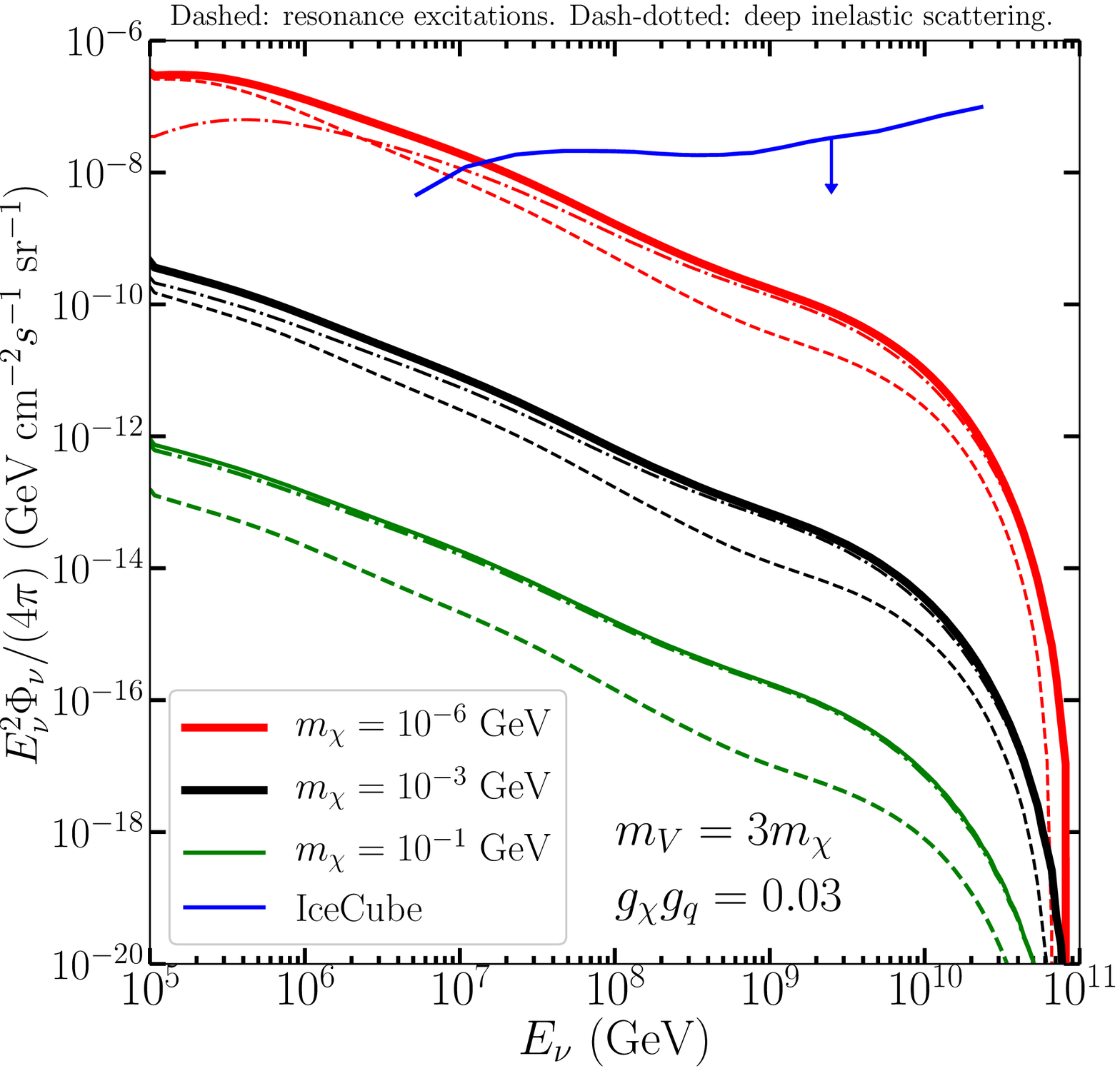}
}
\caption{
The neutrino fluxes summing over all flavors resulting from $\chi$--$p$ inelastic scattering 
for $m_V=1$ eV [panel (a)] and $m_{V}=3 m_\chi$ [panel (b)]. The existing upper limits from IceCube are also shown (blue curves). 
}
\label{fig:nu_flux}
\end{centering}
\end{figure*}

\subsection{Constraints from IceCube}\label{sec:cons-nu}

Both the accelerated DM and the secondary neutrinos can be detected by 
IceCube.
As the DM signals are almost indistinguishable from the neutral-current neutrino events \cite{Guo:2020drq},
a combined analysis of these signals should, in principle, be carried out. 
In this study, however, we choose to do separate analyses for HE DM and neutrino signals. As shown below, the secondary neutrinos can set stronger limits than the DM signals for most of the DM masses considered.        

We first consider the secondary neutrino signals. Fig.~\ref{fig:nu_flux} shows the resulting neutrino fluxes summing over all flavors from resonance excitations and DIS for $m_\chi=10^{-6}$, $10^{-3}$, and $10^{-1}$ GeV, respectively. As discussed in Sec.~\ref{sec:secondaries}, we take the isotropic approximation and show in the figure the averaged flux per solid angle, $\Phi_\nu(E_\nu)/(4\pi) = 1/(4\pi) \times \int d\Omega \phi_\nu(E_\nu, \Omega)$, with $\phi_\nu(E_\nu, \Omega)$ given in Eq.~\eqref{eq:gammanu_flx}. For $E_\nu \gtrsim$ PeV, the fluxes are mainly from DIS. With $E_\nu \propto E_p$, $\phi_p(E_p) \propto E_p^{-\gamma}$, and $\sigma_{\rm DIS} \propto \widetilde E_\chi^{1/3}$, the neutrino spectra scale as $\phi_p(E_p)\sigma_{\rm DIS}(\widetilde E_\chi) \propto  E_\nu^{-\gamma+1/3}$. With $\gamma\approx 3$ and 3.3 below and above the ankle followed by a sharp drop of the CR flux at $E_p \gtrsim 3\times 10^{10}$ GeV due to the GZK cutoff, the neutrino spectra can be quantitatively understood.
Meanwhile, the neutrino flux decreases as $m_\chi^{-1}$. To constrain the DM model, the existing limits on the diffuse fluxes of HE neutrinos from IceCube \cite{Aartsen:2016ngq,Kopper:2017zzm,Aartsen:2018vtx} and Auger \cite{Aab:2015kma} as well as ANITA \cite{Allison:2018cxu} can be directly used. In our study, we use the relevant limits at the 90\% C.L. from Ref.~\cite{Kopper:2017zzm} for $5\times 10^6\lesssim E_\nu \lesssim 3\times 10^7$ GeV and Ref.~\cite{Aartsen:2018vtx} for $3\times 10^7 \lesssim E_\nu \lesssim 8\times 10^{10}$ GeV as shown in Fig.~\ref{fig:nu_flux}.
Since $E_\nu^2 \Phi_\nu$ decreases with $E_\nu$ while the IceCube upper limit increases,
the bounds on the couplings, $g_\chi g_q$, are, in fact, determined by comparing the secondary neutrino flux and the upper limit from the IceCube publications at $E_\nu=5$ PeV.
Fig.~\ref{fig:limits} shows the exclusion limits for $g_\chi g_q$ as functions of $m_\chi$.
Since the flux decreases with $m_\chi$ approximately as $m_\chi^{-1}$, the upper limit on $g_\chi g_q$ increases as $\sim$$m_\chi^{1/2}$. We notice that IceCube has accumulated around 100 events with deposited energies above $\sim$10 TeV, corresponding to an observed all-flavor flux of $E_\nu^2 \phi_{\rm ob} \approx 2 \times 10^{-8}~{\rm GeV~cm^{-2}~s^{-1}~sr^{-1}}$ for $0.3 \lesssim E_\nu \lesssim 3$ PeV (see Fig.~2 of Ref.~\cite{Kopper:2017zzm}). Taking into account the uncertainties for the observed data,
extending the energy window to $E_\nu \lesssim 1$ PeV
does not help to strengthen the bounds on the coupling constants.

The situation for DM signals is more complicated, since the related detector responses are not clear. Considering that no event with deposited energy above 5 PeV has been observed, we simply require the expected number $n_{\chi}$ of DM-induced events at IceCube with deposited energies above 5 PeV to be smaller than 2.4, which leads to the Feldman-Cousins upper limit at the 90\% C.L. for the case of negligible background \cite{Feldman:1997qc}. Considering the potential contributions from the cosmogenic HE neutrinos, the limit derived this way is a conservative one. The expected number $n_\chi$ can be estimated as 
\begin{align}
n_\chi \approx \int d\Omega \int_{5~\rm PeV} d\nu ~ M_TN_AT \Phi_{\chi,~\rm att}^{\rm MW}(T_\chi, \Omega) \frac{d\sigma_{\rm DIS}}{d\nu}\;,        
\end{align}
where $M_T \approx 400$ Mton is the IceCube effective mass at HE regions \cite{Aartsen:2013jdh}, $N_A$ is the Avogadro constant, $T\approx$ 3142.5 days is the exposure time taken from Ref.~\cite{Aartsen:2018vtx}, and $\Phi_{\chi,~\rm att}^{\rm MW}$ is the attenuated DM flux arriving at the detector located 1450 meters below the surface. For 
detecting the HE DM at IceCube, we need only to consider the scattering cross section from DIS   
\begin{align}
 \frac{d\sigma_{\rm DIS}}{d\nu} = \int dQ^2 \frac{d^2\sigma_{\rm DIS}}{d\nu dQ^2}\;.  
\end{align}
To obtain $\Phi_{\chi,~\rm att}^{\rm MW}$, we take into account the energy loss of HE DM when passing through Earth/ice-shell from different declinations and   
follow Ref.~\cite{Bringmann:2018cvk} to treat the attenuation effects. Again, DIS is the dominant channel to be considered for the energy losses of HE DM. The energy loss rate per unit distance is given by 
\begin{align}
 \frac{dT_\chi}{dx} = - n_N \int\frac{d\sigma_{\rm DIS}}{d\nu}\nu d\nu \;,  
\end{align}
where $n_N$ is the number density of nucleons. For DM arriving from different declinations, the lengths traversed by DM inside the Earth are different. We use the Earth density profile from Ref.~\cite{McDonough:2003} for our study.

The excluded region on $g_\chi g_q$ by considering the DM signals at IceCube has been shown in Fig.~\ref{fig:limits} (see the gray region). Unlike the limits from the secondary $\gamma$-rays and neutrinos, the excluded region from DM signals has an upper bound due to the Earth attenuation effects. For $m_\chi \gtrsim 10^{-7}$ GeV, the bounds from the upscattered DM are weaker than those from the secondary neutrinos. The reason is that the DM flux decreases rapidly with $T_\chi$ and is much smaller than that of the secondary neutrinos at high energies above $\sim$PeV.
Without considering the attenuation effects, the expected number of DM signals is proportional to $\Phi_\chi \sigma_{\rm DIS} \propto m_\chi^{-2} (g_\chi g_q)^4$, which is 
different from the secondary flux that scales as $(g_\chi g_q)^2$. Consequently, the lower limit of $g_\chi g_q$ would grow as $m_\chi^{1/2}$. However, this simple argument is no longer valid when the attenuation effect is taken into account. Relevant to the lower limits, the expected number of DM events grows more slowly than $(g_\chi g_q)^4$. This finally results in a more rapid increase of the lower bound of $g_\chi g_q$ approximately as $m_\chi$ and a limited exclusion region. As discussed above, due to a rapidly decreasing differential cross section $d\sigma_{\rm DIS}/dQ^2$, the upscattered DM fluxes at high energies are suppressed. Therefore, compared to Ref.~\cite{Guo:2020drq}, which assumed a constant cross section and differential cross section, the exclusion region derived in this work shrinks largely.

\subsection{Constraints from low-energy neutrino and DM experiments}\label{sec:cons-dm} 

\begin{figure*}[htbp]
\begin{centering}
\subfloat[]{
\includegraphics[width=0.51\textwidth]{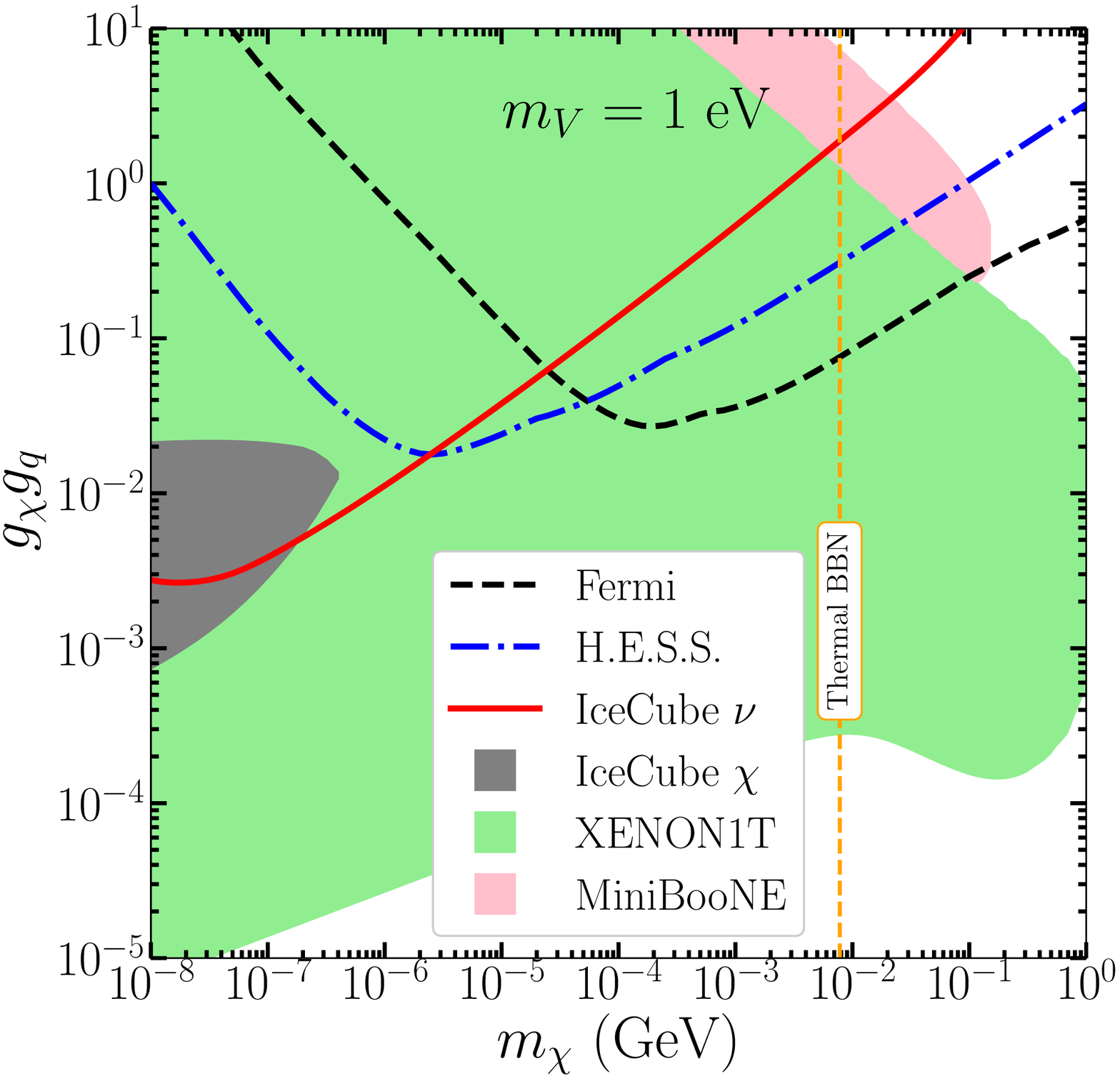}
}
\subfloat[]{
\includegraphics[width=0.51\textwidth]{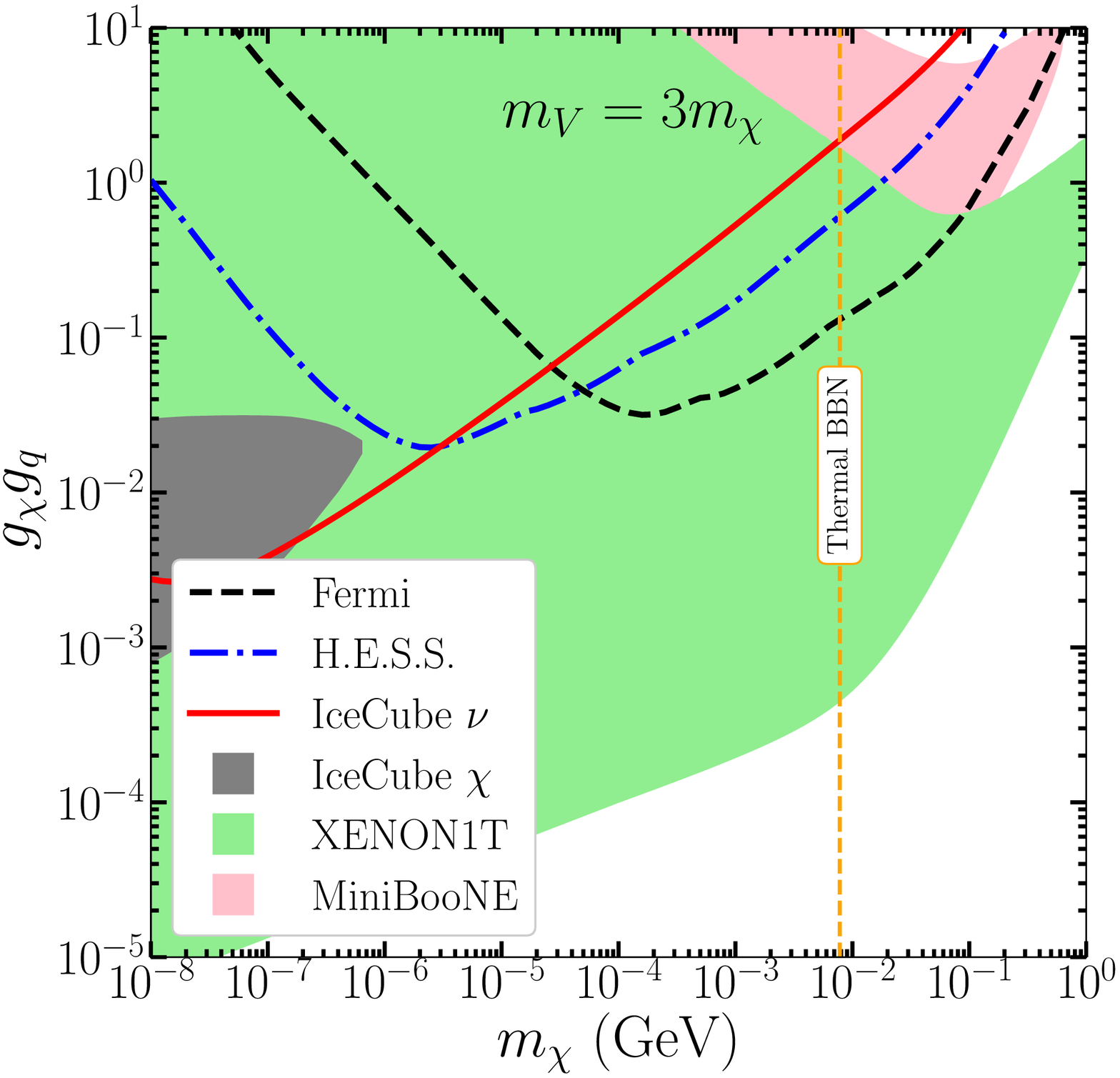}
}
\caption{
The summary of different constraints for $m_V=1$ eV [panel (a)] and $m_{V}=3 m_\chi$ [panel (b)].
The shaded regions are the exclusions by considering the DM signals at XENON1T (green), MiniBooNE (pink), and IceCube (gray).
The black dashed line, blue dash-dotted line, and red solid line are the upper limits derived based on the $\gamma$-ray data from
Fermi, H.E.S.S., and the IceCube HE $\nu$ data, respectively. The BBN bound $m_\chi \lesssim 7.8$ MeV for a Dirac fermion DM \cite{Krnjaic:2019dzc} is also shown.
}
\label{fig:limits}
\end{centering}
\end{figure*}

For the vector portal model considered in this work, the differential cross section of $\chi$-$p$ scattering at low $Q^2$ is dominated by the elastic channel and is orders of magnitude higher than that at high $Q^2$ (see Fig.~\ref{fig:DM_dxs}). Therefore, as shown in Fig.~\ref{fig:DM_flx}, we expect a much higher DM flux at lower energies that can be detected at low-energy neutrino or DM detectors, resulting in a strong bound on the tested model. Using the subroutines implemented in DarkSUSY \cite{Bondarenko:2019vrb,Bringmann:2018lay}, we show in Fig.~\ref{fig:limits} the constraints on $g_\chi g_q$ from XENON1T (green regions) and MiniBooNE (pink regions), respectively. Only the elastic scattering has been considered for calculating the boosted DM flux and for estimating the event rates at detectors. Note that, for the case with a massless mediator, the energy loss rate of DM passing through the Earth, $dE_\chi/dx \propto \int_0^{T_A^{\rm max}} dT_A ~d\sigma_{\rm el}/dQ^2$ with $Q^2=2m_AT_A$, $m_A=Am_N$ the mass, and $T_A$ the recoil energy of nuclei with atomic mass number $A$, is logarithmically divergent.
This is the reason why we choose $m_V=1$ eV instead of a massless mediator, see also footnote \ref{ft1}. 
We also point out that in the original DarkSUSY subroutines a low-energy cutoff $T_A^{\rm min}=1$ eV has been taken, corresponding to a lower cutoff on $Q^2$ of $(2\times 10^9 A)~{\rm eV}^2$, which is much larger than $m_V^2$ for our light mediator case. Thus, when calculating the DM attenuation, we lower $T_A^{\rm min}$ so that the related $Q^2_{\rm min}$ is far below $m_V^2$. We have checked explicitly that our results are insensitive to the cutoff.

The searches for CR-boosted DM in XENON1T set the most stringent bounds for most of the DM mass range explored in this work due to the significantly enhanced cross section and flux at low energies. With a shallow location and, thus, a small Earth attenuation effect, the large couplings which cannot be probed by deep underground
experiments can be further excluded by the MiniBooNE data (see the pink regions). For the case of $m_V = 3m_\chi$,
the limits on $g_\chi g_q$ are quantitatively similar to those for the light mediator case with $m_V=1$ eV at $m_\chi \lesssim 10^{-2}$ GeV. This is simply because the typical values of $Q^2$ encountered in upscattering the DM by CRs or detecting the DM at detectors are higher than $m_V^2$ for both cases.
However, as $m_\chi$ increases, $m_V^2$
becomes comparable to the typical values of $Q^2$ for the case of $m_V=3m_\chi$, leading to suppressed cross sections and enhanced upper/lower limits on the coupling constants compared to the light mediator case.       

For the vector portal model considered, the bounds set by the secondary signals are typically weaker than those from XENON1T. However, unlike the DM signals at terrestrial detectors, the secondary signals do not suffer from the attenuation effect and
can be used to exclude 
all the strong-coupling regions above the upper limits. Especially for $m_\chi \gtrsim 10^{-2}$ GeV, 
the Fermi data exclude the regions of $g_\chi g_q \gtrsim 0.1$ for the case of a light mediator which cannot be explored 
via the underground DM
experiments. For the regions of coupling constant explored, the associated cross sections between nucleons and DM are high enough so that light DM are thermally populated before BBN \cite{Krnjaic:2019dzc}. Therefore, the observed D/H abundance ratio excludes $m_\chi \lesssim 7.8$ MeV for the case of a Dirac fermion DM \cite{Krnjaic:2019dzc} (see the vertical lines in Fig.~\ref{fig:limits}). See also other tighter constraints from beam-dump experiments \cite{Batell:2014yra,Aguilar-Arevalo:2017mqx,Aguilar-Arevalo:2018wea}, supernova cooling \cite{Rrapaj:2015wgs}, low energy neutron scattering \cite{Barbieri:1975xy}, and rare meson decays \cite{Dror:2017nsg}.

For the sake of comparison to studies which set bounds on $\chi$-$p$ scattering cross section, we also provide some benchmark values of the elastic cross section relevant to the direct detection for given $m_\chi$, $m_V$, and $g_\chi g_q$. Since the scattering cross sections  
depend on the energy scales involved, the non-relativistic limit of the scattering cross section $\sigma_{\chi p}^0 \approx g_B^2g_\chi^2\mu_{\chi p}^2/(\pi m_V^4) \propto \int dQ^2 d\sigma_{\rm el}(Q^2=0)/dQ^2$ with $\mu_{\chi p}$ the reduced mass is often used as a reference for cross section comparison \cite{Batell:2014yra,Dent:2019krz}. For $m_\chi \lesssim 1$ GeV, $\sigma_{\chi p}^0 \sim g_q^2g_\chi^2 (m_\chi/{\rm MeV})^2 ({\rm MeV}/m_V)^4 \times 10^{-25}~{\rm cm^2}$, where we have used the relation $g_B=3g_q$. 
However, we note that 
$\sigma_{\chi p}^0$ is originally introduced for non-relativistic heavy DM and cannot  
be readily applied to DM direct detection for light DM/mediator. 
For this reason, we introduce an effective cross section for $\chi$-$p$ scattering as 
\begin{equation}
\sigma_{\chi p}^{\rm eff} \equiv \int_{Q^2_a}^{Q^2_b} dQ^2 d\sigma_{\rm el}/dQ^2 \sim g_B^2 g_\chi^2 \bar Q^2/[\pi(m_V^2+\bar Q^2)^2],
\end{equation}
where $Q^2_{a,b}$ are the lower/upper bounds of $Q^2$ determined by the detection energy window for nuclear recoils and $\bar Q^2 \sim Q^2_{a,b}$ is the typical (or averaged) value. The effective cross section should depend on $\widetilde E_\chi$. Here, we simply assume that $\widetilde E_\chi$ is large enough to produce a detectable recoil energy in the detectors. Taking xenon as the target and a typical value of $10$~keV for nuclear recoil, $\bar Q^2 \sim (0.05~{\rm GeV})^2$. For $m_V^2 \ll \bar Q^2$, $\sigma_{\chi p}^{\rm eff} \sim g_q^2 g_\chi^2 [(0.05~{\rm GeV})^2/\bar Q^2] \times 10^{-29}~{\rm cm^2}$, and for $m_V^2 \gtrsim \bar Q^2$, $\sigma_{\chi p}^{\rm eff} \sim g_q^2g_\chi^2 [(0.05~{\rm GeV})^2/\bar Q^2] ({\rm GeV}/m_V)^4 \times 10^{-34}~{\rm cm^2}$.

\section{Conclusion and prospect}
\label{sec:conclusion}

As an extension to previous studies, we have investigated the inelastic scattering of CRs on sub-GeV light DM within the Milky Way. At energies high enough to excite the nucleonic resonance states (resonance excitations) or to probe the quark constituent (DIS), the secondary $\gamma$-rays or neutrinos can be produced from hadronization and the subsequent meson decay, and act as another indirect probe of DM. To demonstrate these signatures, we consider a simple vector portal DM model in this work.
We have calculated the expected fluxes of these secondary signals and used the $\gamma$-ray and neutrino telescopes such as Fermi, H.E.S.S., and IceCube to derive limits for
this specific model. On the other hand, the upscattered DM from both elastic and inelastic scatterings are also consistently considered and can be directly detected at low- and high-energy detectors, which can also be used to constrain the same DM model. Because of a much larger cross section of elastic scattering at low energies for this specific model, the low-energy experiments such as XENON1T set stronger bounds than those obtained from the secondary signals and the HE component of the accelerated DM.                                    

The bounds derived are weakly dependent on
the spatial distribution of the CRs and the DM halo profiles (see Fig.~\ref{fig:relative_flx}). 
We have considered only CR protons in our study and assumed that a fraction of 30\% are protons in the HE CRs above $\sim$PeV. We do not expect our results to be affected too much by this assumption
unless the proton abundance in HE CRs is extremely small. To calculate the secondary neutrinos from DIS, we have relied on the SOPHIA code using the Lund Monte Carlo generator JETSET to treat hadronization. Since all hadronization models are relatively well constrained by data and can give rise to quantitatively similar yields even at very HE regions,
the associated uncertainties for the secondary HE neutrino fluxes should not largely affect our conclusion either.                       

Although considering the secondary signals does not %\mkgreen{significantly}\cmtgg{I guess it's not fair to add `significantly' here} 
improve the bounds for the vector portal model for $m_\chi\lesssim 10^{-2}$~GeV, they could be interesting for other DM models with relatively suppressed/enhanced cross sections at low/high energies. One possible scenario is that the DM candidate carries no charge but only couples to the vector mediator via a dipole form; i.e., DM particles have dipole moments. Other possibilities include composite DM scenario or the scenarios that scalar/vector DM particles scatter with nucleons in the $s$-channel. In these cases, the HE signals could possibly lead to a better sensitivity than the low-energy ones. In this work, we choose to take the simple vector portal DM model as a starting point and show the possible relevance of inelastic channels in HE regions. We plan to extend these studies to other interesting models in the future.      

%\newpage

\section*{Acknowledgments}
We thank Hsiang-nan Li for helpful discussions about this project. G.~G. and M.-R.~W. acknowledge support from the Academia Sinica by Grant No.~AS-CDA-109-M11. Y.-L.~S.~T.
was funded in part by the Taiwan Young Talent Programme of Chinese Academy of Sciences 
under the Grant No.~2018TW2JA0005 and 
the Ministry of Science and Technology, Taiwan under the Grant No.~109-2112-M-007-022-MY3.
M.-R.~W. acknowledges support from the Ministry of Science and Technology, Taiwan under Grants No.~108-2112-M-001-010 and No. 109-2112-M-001-004, and the Physics Division, National Center of Theoretical Science of Taiwan.
Q.~Y. is supported by the National Natural Science Foundation of China (No. 11722328 and No. 11851305), 
the 100 Talents Program of Chinese Academy of Sciences and the Program for Innovative Talents and
Entrepreneur in Jiangsu.

\end{document}